\documentclass{emulateapj}
\usepackage{graphicx}

\newcommand\bmath[1] {\mbox{\boldmath$\rm #1$}}

\usepackage[usenames,dvips]{color}

\usepackage{natbib}
\newcommand\s{{\rm s}} 
\newcommand\yr{{\rm yr}} 
\newcommand\GHz{{\rm GHz}} 
\newcommand\m{{\rm m}} 
\newcommand\mm{{\rm m}\m} 
\newcommand\cm{{\rm c}\m} 
\newcommand\pc{{\rm pc}} 
\newcommand\kpc{{\rm k}\pc} 
\newcommand\erg{{\rm erg}} 
\newcommand\mas{{\rm mas}} 
\newcommand\muas{\mu{\rm as}} 

\renewcommand\d{{\rm d}}
\newcommand\e{{\rm e}}
\newcommand\erf{{\rm erf}}

\def\Jy{{\rm Jy}}
\def\Ms{{M_\odot}}
\def\Rs{{r_{\rm S}}}

\def\CARMA{{CARMA}}
\def\JCMT{{JCMT}}
\def\SMTO{{SMTO}}

\begin{document}

\title{Estimating the Parameters of Sgr A*'s Accretion Flow Via Millimeter VLBI}

\author{
Avery E.~Broderick\altaffilmark{1},
Vincent L.~Fish\altaffilmark{2},
Sheperd S.~Doeleman\altaffilmark{2} \&
Abraham Loeb\altaffilmark{3}
}
\altaffiltext{1}{Canadian Institute for Theoretical Astrophysics, 60 St.~George St., Toronto, ON M5S 3H8, Canada; aeb@cita.utoronto.ca}
\altaffiltext{2}{Massachusetts Institute of Technology, Haystack Observatory, Route 40, Westford, MA 01886.}
\altaffiltext{3}{Institute for Theory and Computation, Harvard University, Center for Astrophysics, 60 Garden St., Cambridge, MA 02138.}

\shorttitle{VLBI Parameter Estimation for Sgr A*}
\shortauthors{Broderick et al.}

\begin{abstract}
Recent millimeter-VLBI observations of Sagittarius A* (Sgr A*) have,
for the first time, directly probed distances comparable to the
horizon scale of a black hole.  This provides unprecedented access to the
environment immediately around the horizon of an accreting black
hole.  We leverage both existing spectral and polarization
measurements and our present understanding of accretion theory to
produce a suite of generic radiatively inefficient accretion flow
(RIAF) models of Sgr A*, which we then fit to these recent millimeter-VLBI
observations.  We find that {\em if} the accretion flow onto Sgr A* is
well described by a RIAF model, the orientation and magnitude of the
black hole's spin is constrained to a two-dimensional surface in the
spin, inclination, position angle parameter space.  For each of these
we find the likeliest values and their 1-$\sigma$ \& 2-$\sigma$ errors
to be $a=0^{+0.4+0.7}$,
$\theta={50^\circ}^{+10^\circ +30^\circ}_{-10^\circ-10^\circ}$,
and 
$\xi={-20^\circ}^{+31^\circ+107^\circ}_{-16^\circ-29^\circ}$, when the
resulting probability distribution is marginalized over the others.
The most probable combination is $a=0^{+0.2+0.4}$,
$\theta={90^\circ}_{-40^\circ-50^\circ}$ and
$\xi={-14^\circ}^{+7^\circ+11^\circ}_{-7^\circ-11^\circ}$, though the
uncertainties on these are very strongly correlated, and high
probability configurations exist for a variety of inclination angles
above $30^\circ$ and spins below $0.99$.  Nevertheless, this
demonstrates the ability millimeter-VLBI observations, even with only
a few stations, to significantly constrain the properties of Sgr A*.
\end{abstract}

\keywords{black hole physics --- Galaxy: center --- techniques: interferometric --- submillimeter --- accretion, accretion disks}

\maketitle

\section{Introduction}

Understanding the structure and dynamics of black hole accretion flows
has remained a central problem in astrophysics.  Only in the past
decade have sufficient numerical resources existed to perform
self-consistent, three-dimensional magneto-hydrodynamic (MHD)
simulations capable of resolving the magneto-rotational instability,
believed to be responsible for angular-momentum transport in
black-hole accretion flows.  However, a true {\em ab initio}
computation is still well beyond reach, resulting in the need for a
variety of simplifying assumptions (e.g., the suitability of MHD,
importance of electron-ion coupling, properties of accelerated
electrons, etc.). As a result, the number of applicable models has
rapidly proliferated, many of which are capable of describing the
variety of phenomena observed.   This is due in large part to the
inability of current observations to resolve horizon scales.
Unfortunately, the compact nature of black holes makes it very
difficult to access the inner-edge of black hole accretion flows.

Sagittarius A* (Sgr A*), the bright radio point source coincident with
the center of the Milky Way, is presently the best studied known
black-hole candidate.  Observations of orbiting OB-stars, the closest
of which passes within $45\,{\rm AU}$ of Sgr A*, have produced a
measured mass of $4.5\pm0.4\times10^6\,\Ms$ and an Earth-Sgr A*
distance\footnote{The mass and distance measurements are
  strongly correlated, with mass scaling roughly as $M\propto
  D^{1.8}$.} of $8.4\pm0.4\,\kpc$ \citep{ghez09,gillessen08}.  Already, there is
strong evidence for the existence of horizon in this
source \citep{Brod-Nara:06,Brod-Loeb-Nara:08}.  However, in many ways,
Sgr A* is very different than the supermassive black holes in
AGN. Unlike its active brethren, Sgr A* is vastly under-luminous, with
a luminosity many orders of magnitude smaller than the Eddington
luminosity.  As a result, it is widely expected that Sgr A*'s
accretion flow is quite different from those in AGN, though
perhaps more indicative of the roughly 90\% of supermassive black
holes that are not presently in an active phase.

Even when strong-gravitational lensing is accounted for, the apparent angular
size of Sgr A*'s horizon is only $55\pm2\,\muas$, a factor of two
larger than the next largest black hole (M87) and orders of magnitude
larger than any other known black hole (including all stellar-mass
black holes).  Nevertheless, despite its tiny angular size, recent
millimeter VLBI experiments have successfully resolved this scale
\citep{Doel_etal:08}.  Since only three telescopes were
involved with these observations, the resulting $u$--$v$ coverage of
the measured visibilities is very sparse.  As a result only two simple
models of Sgr A*'s image, a Gaussian and an annulus, were fit to the
data by \citet{Doel_etal:08}.

However, we have a great deal of additional information about Sgr A*,
including its spectral and polarization properties.  We may also
require physical consistency in any model (which would likely rule out
an annulus, for example).  Furthermore, we would like to evaluate the
ability of, and optimize for this purpose, future millimeter and
sub-millimeter VLBI experiments to constrain fundamental properties of
the accretion flow in Sgr A*.  This paper demonstrates the fitting
and parameter estimation procedure for a simple radiatively
inefficient accretion disk model (RIAF).  In particular, we show that
even from very sparse baseline coverage it is possible to robustly
extract interesting parameters of a generic RIAF.  A study that
considers how future high frequency VLBI observations can extend this
work will be presented elsewhere \citep{fish08}.

In the analysis presented here, we make full use of all the
observations described in \citet{Doel_etal:08}.  These include two days
of VLBI observations at 230GHz that targeted Sgr A* using an array
consisting of the James Clerk Maxwell Telescope (\JCMT) on Mauna Kea,
the Arizona Radio Observatory Submillimeter Telescope (\SMTO) on Mt.
Graham in Arizona, and one $10,\m$ dish of the Coordinated Array for
Research in Millimeter-wave Astronomy (\CARMA) in California.   Robust
detections of Sgr A* and correlated flux density measuremnets were
obtained on the \CARMA-\SMTO~and \JCMT-\SMTO~baselines.  No detections
were found on the \CARMA-\JCMT~baseline resulting in upper limits on the
correlated flux density.  In addition, contemporaneous total flux
density measurements at the same frequency were obtained by using the
full \CARMA~array operating as a stand-alone instrument.  Full details
of the observations, calibration, and data processing can be found in
\citet{Doel_etal:08}.

In section \ref{sec:RIAF} we present the simple
accretion flow model that we employ.  Sections \ref{sec:DataAnalysis} and
\ref{sec:Results} discuss the data fitting procedure, including the
Bayesian method by which we do the parameter estimation, and the
generic constraints placed by the current VLBI results.  Section
\ref{sec:parameters} details how we define our uncertainties and
presents our parameter estimates.  Finally, we conclude in section
\ref{sec:Conclusions}.

\section{RIAF Visibility Modeling}\label{sec:RIAF}

\begin{figure}
\begin{center}
\includegraphics[width=\columnwidth]{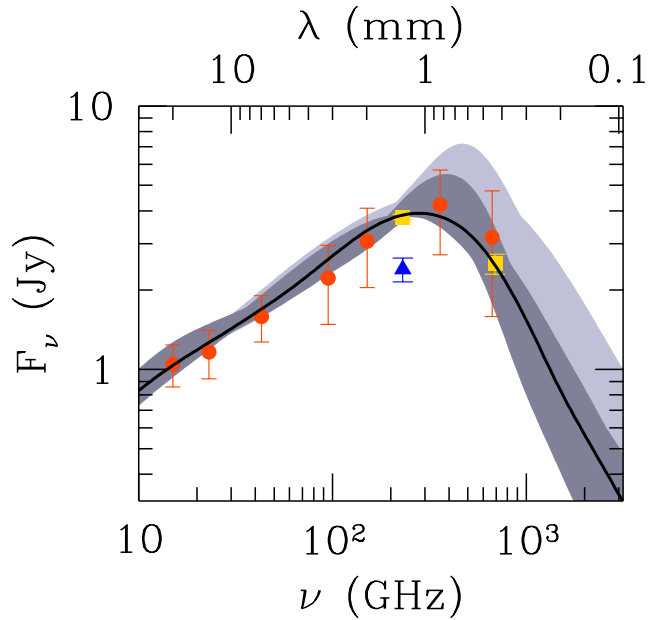}
\end{center}
\caption{High-frequency radio data and the range of fitted model
  spectra.  The red circles are taken from \citet{yuan04} (and
  references therein), the yellow squares are from \citet{marrone06a},
  and the blue triangle is the single-dish flux during the VLBI
  observations discussed here.  Since all of the data points were not
  taken coincidentally errorbars on the radio data shown by the red
  circles are indicative of the variability, not the intrinsic
  measurement errors (in contrast to the yellow squares which were
  determined by coincident observations, and the blue triangle not
  used in the fit).  The gray regions show the envelope of the model
  spectra, with the dark gray region showing $a\le0.9$, and the light
  gray showing $0.9<a\le0.998$.  Finally, the black line shows the
  spectrum for the spin and inclination shown in Fig. \ref{fig:IVV}
}\label{fig:spectra}
\end{figure}

\begin{figure*}[t!]
\begin{center}
\includegraphics[width=0.32\textwidth]{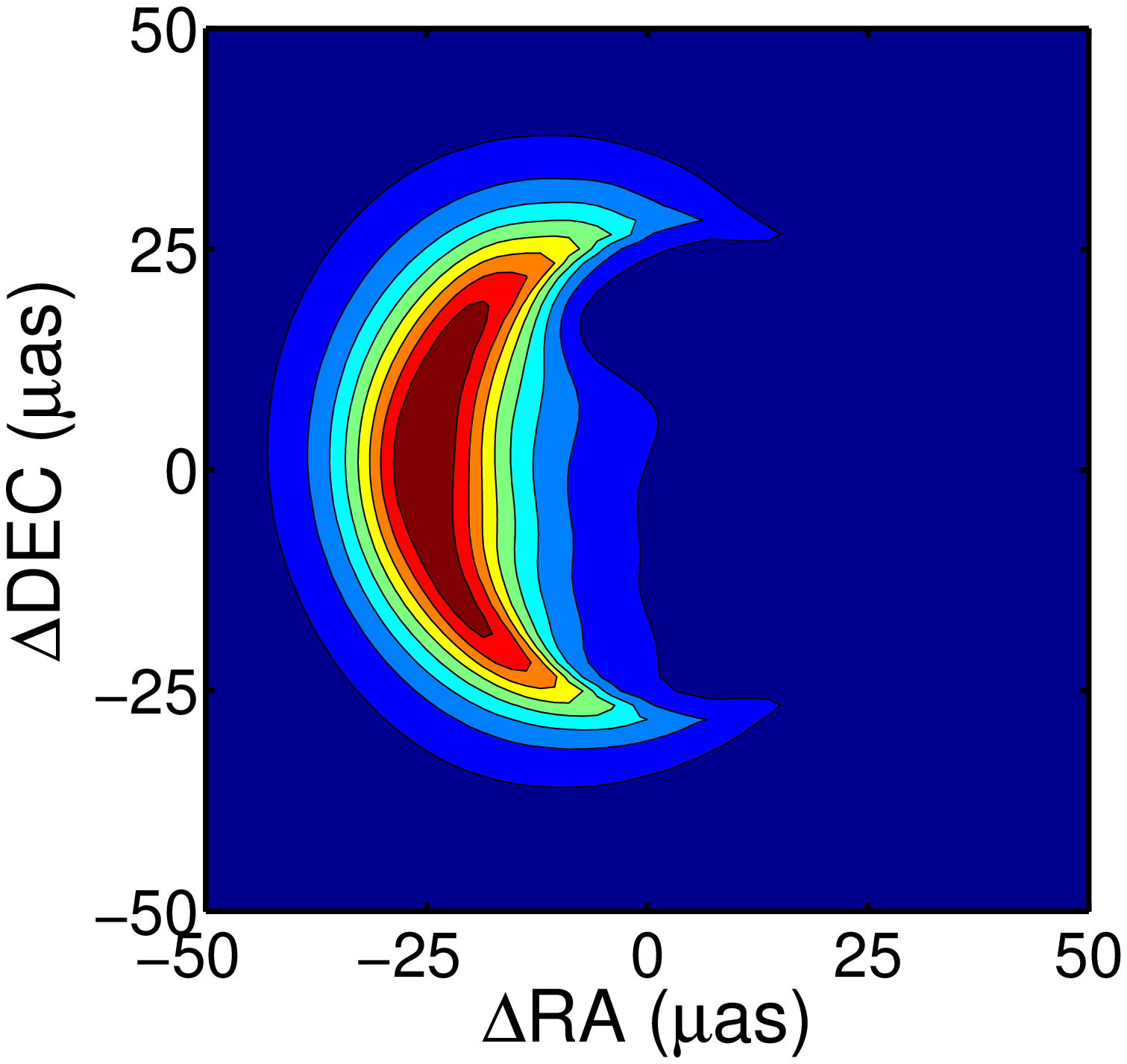}
\includegraphics[width=0.32\textwidth]{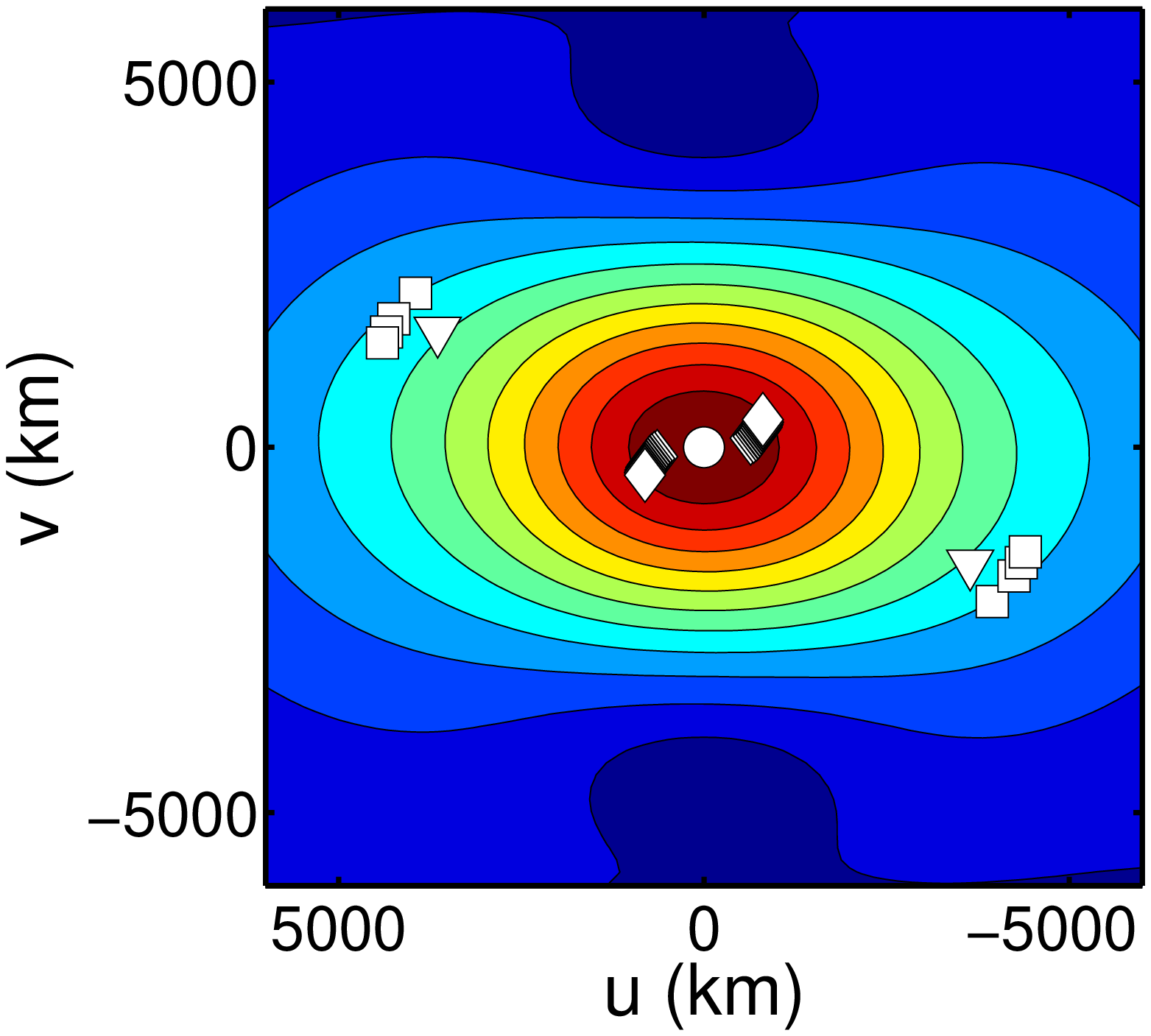}
\includegraphics[width=0.32\textwidth]{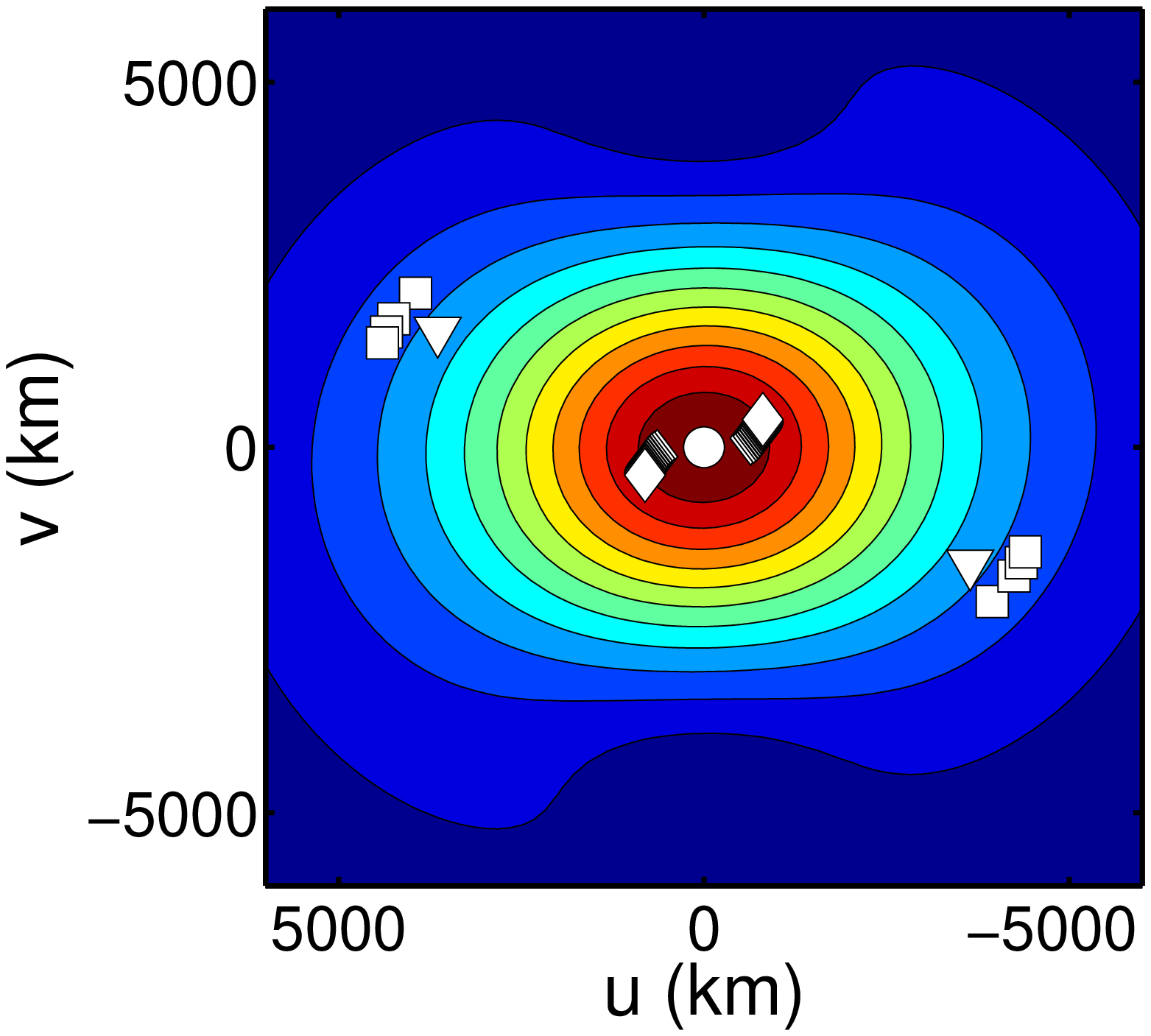}
\end{center}
\caption{{\em Left:} An example image of the radiatively inefficient
  accretion flow around Sgr A* at $230\,\GHz$, centered on the black
  hole, viewed at $60^\circ$
  from the black hole spin axis, which is oriented towards North
  ($\xi=0$), for a moderately rotating ($a=0.5$) black hole.  The
  image appears as an crescent due primarily to the Doppler beaming and
  shifting associated with the relativistic orbital velocities.  The
  color-scheme is normalized such that red is bright, blue corresponds
  to vanishing intensity and the total flux from the image is
  $2.4\,\Jy$.  {\em Middle:} Visibility magnitudes associated with the
  image shown on top.  For reference, the positions of the
  observations in the $u$--$v$ plane are also show by the white points
  (the circle, diamonds, \& squares correspond to the single-dish,
  \SMTO-\CARMA, \& \JCMT-\SMTO~ detections, and the triangle
  corresponds to the \CARMA-\JCMT~upper limit). {\em Right:} Scatter
  broadened visibility magnitudes.  Again the $u$--$v$ positions of
  the VLBI observations are shown.}\label{fig:IVV}
\end{figure*}

\subsection{Accretion Flow Modeling}
Sgr A* transitions from an inverted, optically thick
spectrum to a optically thin spectrum near millimeter wavelengths.  This
implies that Sgr A* is only becoming optically thin at $1.3\,\mm$.
Due to relativistic effects this transition does not occur
isotropically for orbiting gas \citep[e.g.,][]{Brod-Loeb:06a},
becoming optically thin on the receding side at longer wavelengths
than the approaching side of the accretion flows orbit.  As a
consequence, the opacity of  the underlying accretion flow is
crucial to imaging Sgr A*'s accretion flow.

Despite being diminutive in comparison to the Eddington luminosity for
a $4.5\times10^6\,\Ms$ black hole, Sgr~A* is still considerably bright,
emitting a bolometric luminosity of approximately $10^{36}\,\erg/\s$.
Thus, it has been widely accepted that Sgr A* must be
accretion powered, implying a minimum accretion rate of at least
$10^{-10}\,\Ms/\yr$. It is presently unclear how this emission is
produced.  This is evidence by the variety of models that have been
proposed to explain the emission characteristics of Sgr A*
\citep[e.g.,][]{narayan98,blandford99,falcke00a,yuan02,yuan03,loeb07}.
Models in which the emission arises directly from the accreting
material have been subsumed into the general class of RIAFs, defined
by the generally weak
coupling between the electrons, which radiate rapidly, and the ions,
which efficiently convert gravitational potential energy into heat
\citep{narayan98}. This coupling may be sufficiently weak to allow
accretion flows substantially in excess of the $10^{-10}\,\Ms/\yr$
required to explain the observed luminosity with a canonical radiative
efficiency.

Nevertheless, the detection of polarization from Sgr A*
above $100\,\GHz$ \citep{aitken00,bower01,bower03,marrone06}, and
subsequent measurements of the Faraday rotation measure
\citep{macquart06,marrone07b}, has implied that the accretion rate
near the black hole is significantly less than the Bondi rate,
requiring the existence large-scale outflows
\citep{agol00,quataert00}. Therefore, in the absence of an unambiguous
theory, we adopt a simple, self-similar model for the underlying
accretion flow which includes substantial mass-loss.

For concreteness, as in \citet{Brod-Loeb:06a}, we follow
\citet{yuan03}, and employ a model in which the accretion flow has a
Keplerian velocity distribution, a population of thermal electrons
with density and temperature
\begin{equation}
n_{e,\rm th} = n^0_{e,\rm th} \left(\frac{r}{\Rs}\right)^{-1.1} \e^{-z^2/2 \rho^2}
\end{equation}
and
\begin{equation}
T_{e} = T^0_{e} \left(\frac{r}{\Rs}\right)^{-0.84}\,,
\end{equation}
respectively, a population of non-thermal electrons
\begin{equation}
n_{e,\rm nth} = n^0_{e,\rm nth} \left(\frac{r}{\Rs}\right)^{-2.9} \e^{-z^2/2 \rho^2}\,,
\end{equation}
and a toroidal magnetic field in approximate ($\beta=10$)
equipartition with the ions (which produce the majority of the
pressure), i.e.,
\begin{equation}
\frac{B^2}{8\pi}
=
\beta^{-1} n_{e,\rm th} \frac{m_p c^2 \Rs}{12 r}\,.
\end{equation}
In all of these, $r_s=2GM/c^2$ is the Schwarschild radius, $\rho$ is
the cylindrical radius and $z$ is the vertical coordinate.  Inside of
the innermost-stable circular orbit (ISCO) we assume the gas is
plunging upon ballistic trajectories.  In all of these expressions the
radial structure was taken directly from \citet{yuan03} and the
vertical structure was determined by assuming the disk height is
comparable to $\rho$.  Given a choice for the coefficients and a
radiative transfer model, images may then produced using the
fully-relativistic ray-tracing and radiative transfer schemes
described in \citet{Brod-Loeb:06a} and \citet{Brod-Loeb:06b}.

The primary emission mechanism is synchrotron, arising from both the
thermal and non-thermal electrons.  We model the emission from the 
thermal electrons using the emissivity described in \citet{yuan03},
appropriately altered to account for relativistic effects \citep[see,
  e.g.,][]{Brod-Blan:04}.  Since we perform polarized radiative
transfer via the entire complement of Stokes parameters, we employ the
polarization fraction for thermal synchrotron as derived in
\citet{petrosian83}.  In doing so we have implicitly assumed that
the emission due to thermal electrons is isotropic, which while
generally not the case is unlikely to change our results
significantly.  For the non-thermal electrons we follow
\citet{jones77} for a power-law electron distribution, cutting the
electron distribution off below a Lorentz factor of $10^2$ and
corresponding to a spectral index of $\alpha_{\rm disk}=1.25$, both roughly
in agreement with the assumptions in \citet{yuan03}.  For both the
thermal and non-thermal electrons the absorption coefficients are be
determined directly via Kirchoff's law.

As in \citet{Brod-Loeb:06a}, to correct for the fact that
\citet{yuan03} was a Newtonian study, the three coefficients
($n^0_{e,\rm th}$, $T^0_e$ and $n^0_{e,\rm nth}$) were adjusted to fit
the average radio, sub-millimeter and near-infrared spectrum of Sgr
A*.  However, our procedure is different from that employed in
\citet{Brod-Loeb:06a} in two respects.  Firstly, we keep the radial
index of the non-thermal
electrons fixed for all models.  Secondly, the fitting is performed
systematically for a large number of positions in the
inclination--spin parameter space, yielding a tabulated set of
coefficient values.  For every inclination and black
hole spin presented here this was possible with extraordinary accuracy
(reduce $\chi^2<1$ in all cases and $\lesssim0.2$ for many), implying
that this model is presently significantly under-constrained by the
quiescent spectrum alone\footnote{Part of the reason for this
is almost certainly the fact that many of the radio fluxes were
measured during different observational epochs, and thus the flux
uncertainties are indicative of the source variability, not the
intrinsic measurement error.  For two of the data points in
Fig. \ref{fig:spectra} (the yellow squares), this is not the case,
having been measured coincidentally, and thus these play a much more
significant role in constraining the RIAF model parameters.}.  
Subsequently, we obtained the appropriate model parameters for
arbitrary spins and inclinations via a high-order polynomial
interpolation.

The range of spectra for the models we employ here is
presented in Fig. \ref{fig:spectra}, together with the radio \&
sub-millimeter data that was used for the fitting procedure.  The
shown spectra deviate from the fitted spectra much more significantly
than the scatter in the fitted spectra themselves.  This is due almost
entirely to interpolation process, and is most extreme at high-spins,
where the coefficient values begin to change rapidly (and thus have
larger interpolation errors).  Nevertheless, these do not produce
substantially different images at $1.3\,\mm$ from their tabulated
counterparts, and it is important to note that the associated error in
the underlying RIAF model serves only to favor very high spins
($a\ge0.99$), since in these cases it results in a larger
accretion-flow photosphere and, as we shall see, the small size of the
photosphere of the high-spin models is the primary reason they are
excluded.  Hence, in reality, such models are more strongly disfavored
than shown here.  All of these models are also capable of producing
the Faraday rotation  measures observed, and thus the polarimetric
properties of Sgr~A*.

At the time that Sgr A* was being monitored by
\citet{Doel_etal:08} it exhibited an anomalously low $1.3\,\mm$ flux
of $2.4\pm0.25\,\Jy$, almost $40\%$ below its average value (shown by
the blue triangle in Fig. \ref{fig:spectra}).  To appropriately
account for this we chose a ``minimal'' prescription for changing the
model, reducing all densities by a fixed factor (decreasing the
magnetic field such that $\beta$ is fixed) until the observed flux was
reproduced.  In a sense, this models the low flux as a low-mass
accretion period.  These lower-density models do have noticeably
smaller, though qualitatively similar, images compared to those used
in the spectral fits.

An example image is shown in the left-panel of Fig. \ref{fig:IVV} for
a moderate spin and viewing inclination.  As seen in previous efforts
to image relativistic accretion flows, the flow appears as a crescent
associated with the approaching side of the accretion disk
\citep{Brod-Loeb:06a,Brod-Loeb:06b}.  The non-negligible optical depth
obscures the black hole ``silhouette'' on this side.
The receding side is all but invisible as a consequence of the
Doppler shift and Doppler beaming associated with the
relativistic orbital motion in the inner-most portions of the
accretion flow.
Images for $\theta>90^\circ$ are related by a reflection (across a
line perpendicular to the projected spin axis) to images with an
inclination $180^\circ-\theta$.  However, due to the approximate
reflection symmetry of the RIAF model images, the constraints arising
from spins pointed counter to and along the line of sight are nearly
identical.  As a result, for clarity we restrict ourselves to the
latter range.

Using this model we created a library of ideal-resolution images, each
with a flux of $2.4\,\Jy$, having different spins and viewing
inclinations spanning all possible values.

\begin{figure}[t!]
\begin{center}
\includegraphics[width=\columnwidth]{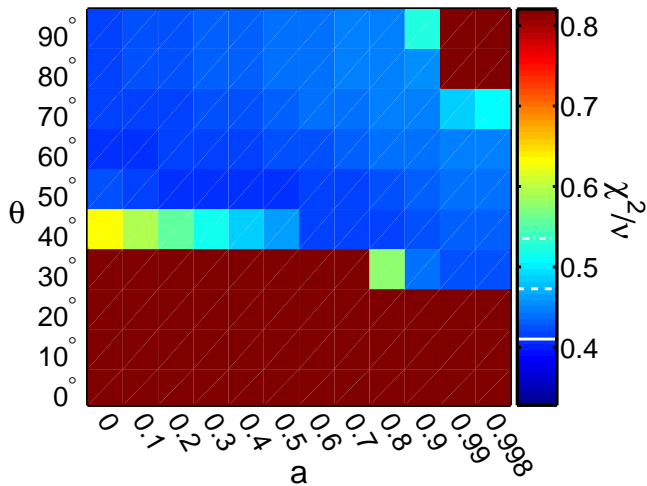}
\end{center}
\caption{Effective reduced $\chi^2$ for the best fit models in $\xi$
  and $V_{0}$ as a function of inclination and spin.  For reference,
  white lines on the color-bar show the minimum $\chi^2$ (solid),
  minimum $\chi^2+1$ (dashed) and minimum $\chi^2+2$ (dot-dashed).
  We were able to find a good fit for nearly all inclinations above
  $40^\circ$.  Models with lower inclinations are too large at all
  $\xi$ to produce the observed flux on the \JCMT-\SMTO~baseline.
  Conversely, models with large inclinations and high spins were
  over-predict the fluxes on the \SMTO-\CARMA~and
  \CARMA-\JCMT~baselines.} \label{fig:csq}
\end{figure}

\begin{figure*}[th!]
\begin{center}
\includegraphics[width=0.49\textwidth]{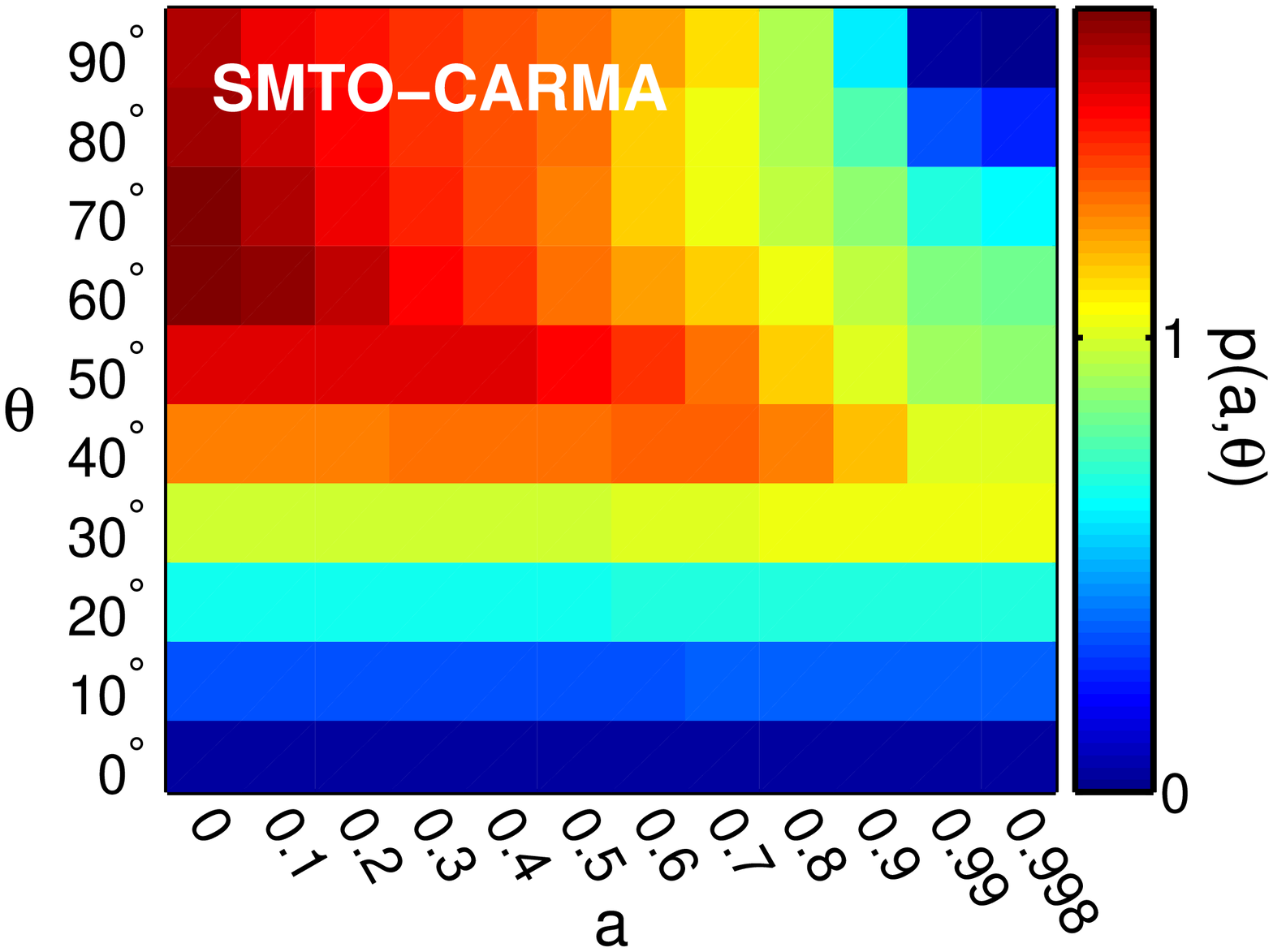}
\includegraphics[width=0.49\textwidth]{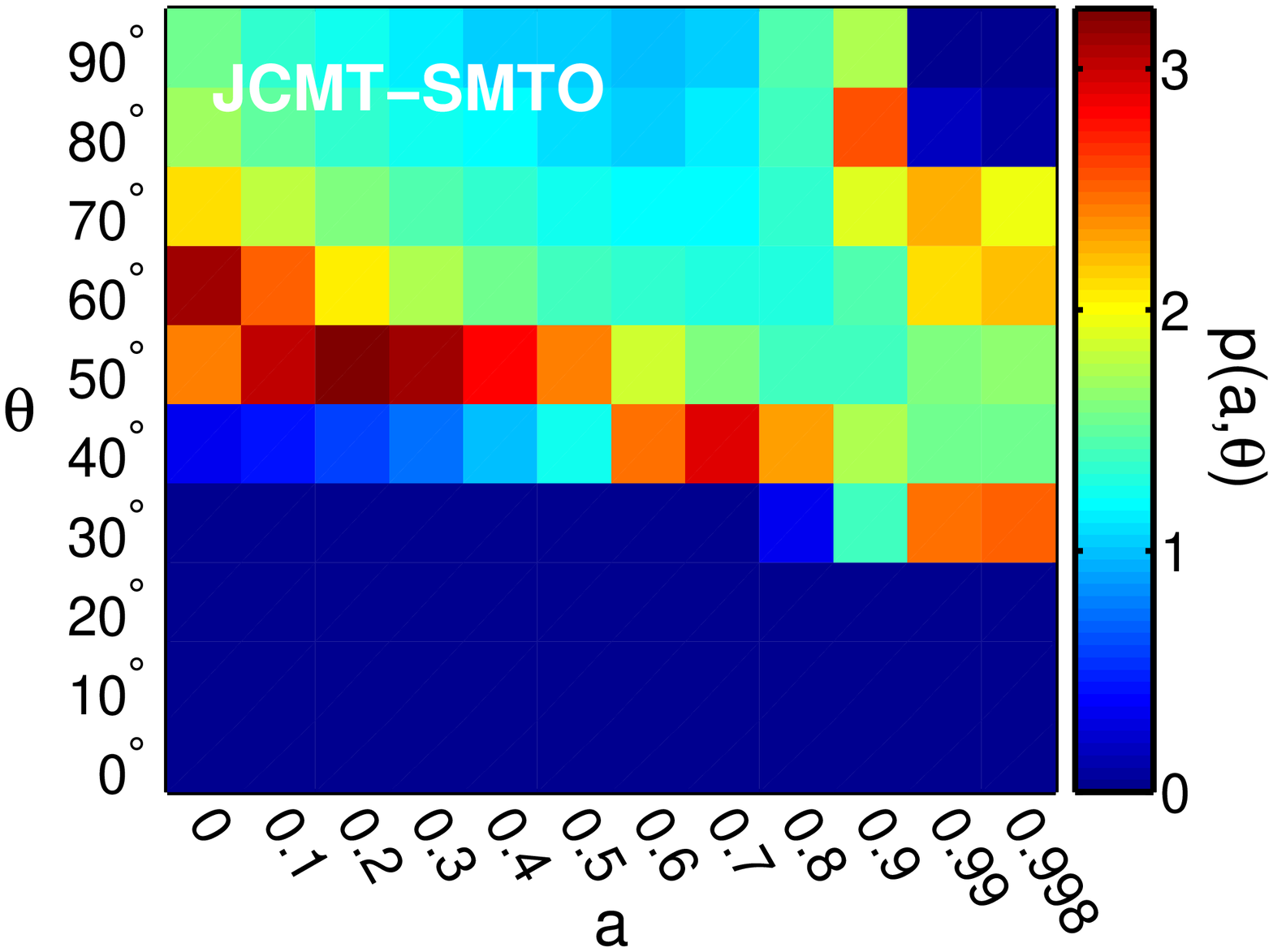}\\
\includegraphics[width=0.49\textwidth]{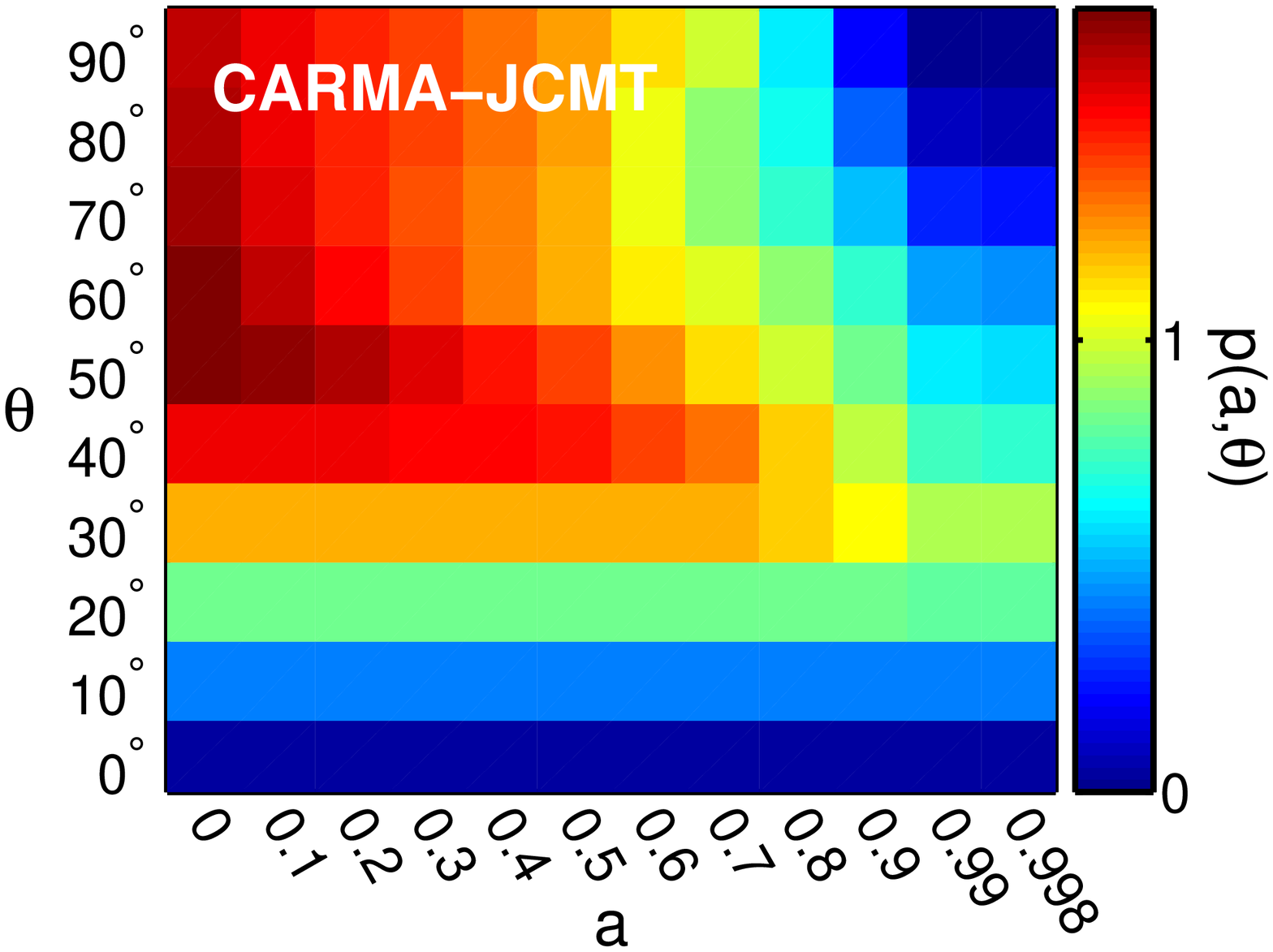}
\includegraphics[width=0.49\textwidth]{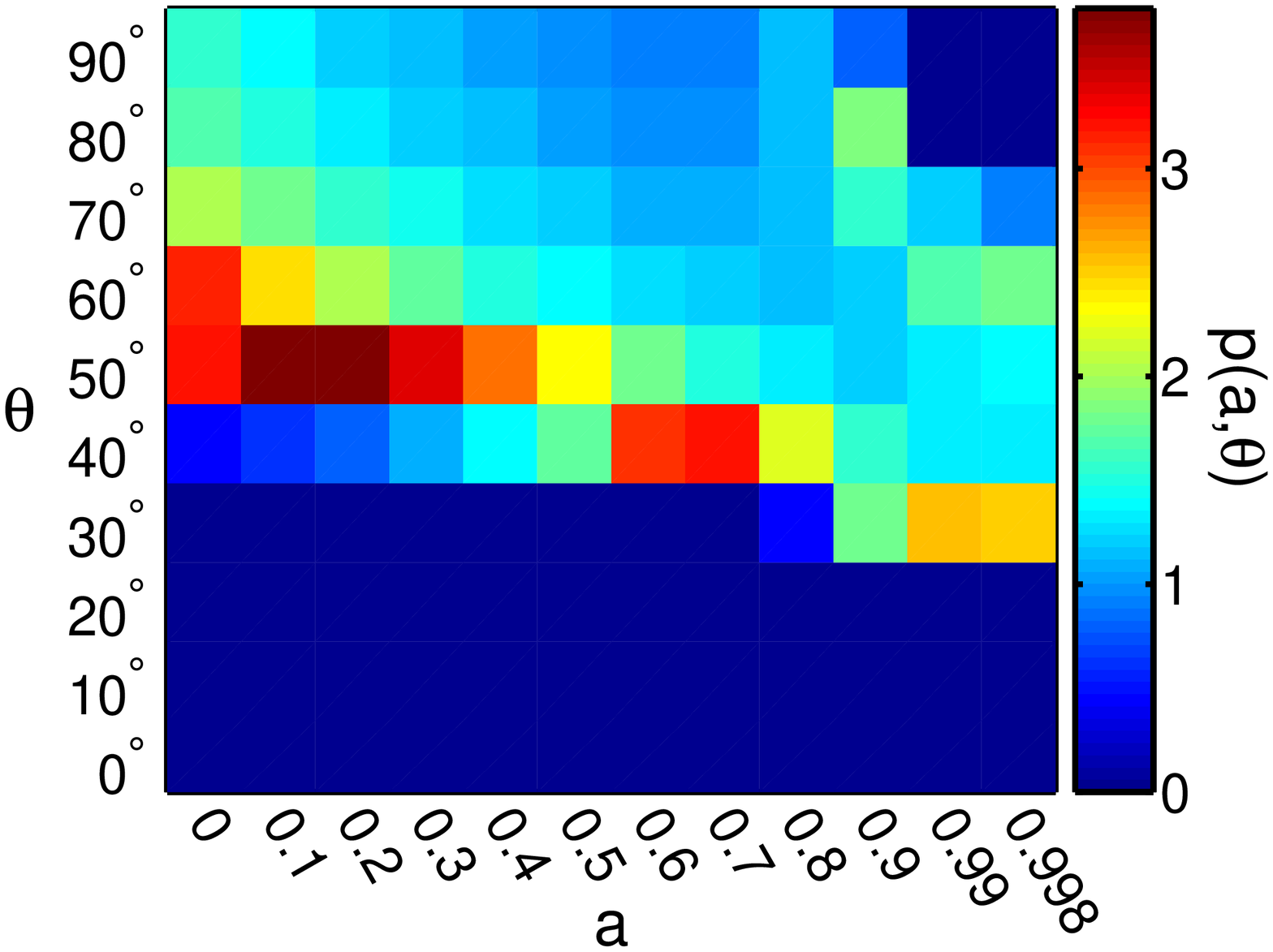}
\end{center}
\caption{Probability densities of a given inclination and black hole
  spin for our canonical Sgr A* model.  The {\em top left}, {\em top
    right} and {\em bottom left}
  panels show the contribution to $p(a,\theta)$ from the \SMTO-\CARMA,
  \JCMT-\SMTO~and \CARMA-\JCMT~baselines, after being marginalized
  over $\xi$ and at the most-likely $V_0$.  The combined implied
  $p(a,\theta)$, marginalized over $\xi$ and $V_0$ is shown in the
  {\em bottom right}.  (Islands of high probability in the right
  panels are an artifact of under sampling the probability peak in
  inclination.) In each, $p(a,\theta)$ is normalized such that the
  average is unity, providing a clear sense of the significance of the
  variations in probability.  Note that the color-scale is different
  in each panel.  While the \JCMT-\SMTO~measurement is
  clearly the most constraining, the other baselines are critical to
  eliminating high-inclination, high-spin solutions.}\label{fig:marg}
\end{figure*}

\subsection{Interstellar Electron Scattering}
The existence of an interstellar scattering screen between Earth and
the Galactic center has been well known for some time now.  This has
been carefully characterized empirically by a number of authors; we
use the recent model from \citet{bower06}.  In this the observed flux
distribution is obtained by convolving the ideal flux with an
anisotropic Gaussian scattering kernel.  The anisotropic Gaussian is
defined by the scattering widths along major and minor axes
(both $\propto\lambda^2$) and the position angle of the minor axis
(which is independent of $\lambda$).  From \citet{bower06}, the
associated full-width, half-max for the major and minor axes are
\begin{eqnarray}
{\rm FWHM}_{\rm maj} &= \displaystyle 1.309 \left(\frac{\lambda}{1\,\cm}\right)^2\,\mas\\
{\rm FWHM}_{\rm min} &= \displaystyle 0.64 \left(\frac{\lambda}{1\,\cm}\right)^2\,\mas
\end{eqnarray}
with the major axis oriented $78^\circ$ East of North.  In practice,
the broadening was done in the $u$--$v$ plane, where it reduces to a
multiplicative factor.

\begin{figure*}
\begin{center}
\includegraphics[width=0.32\textwidth]{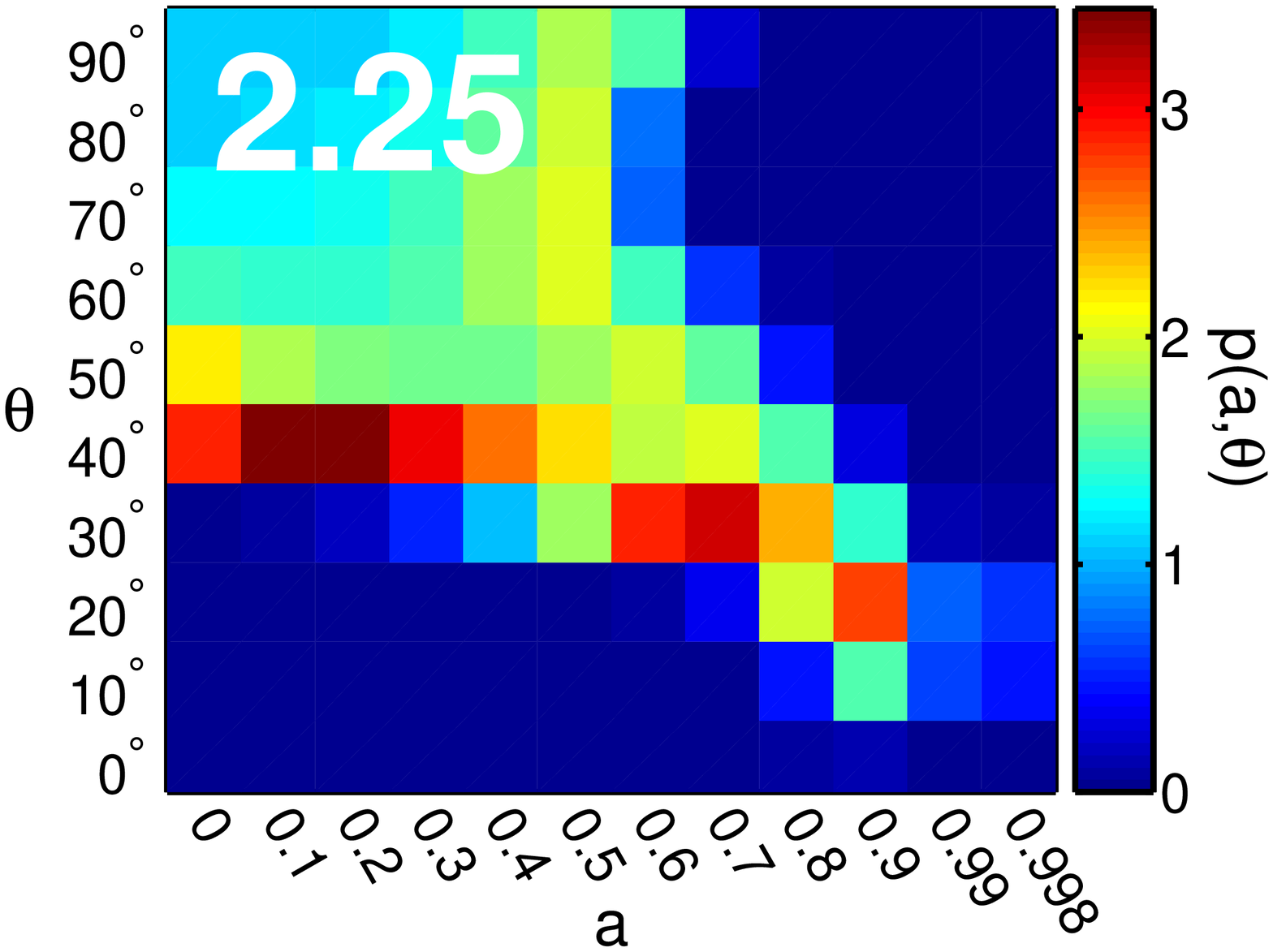}
\includegraphics[width=0.32\textwidth]{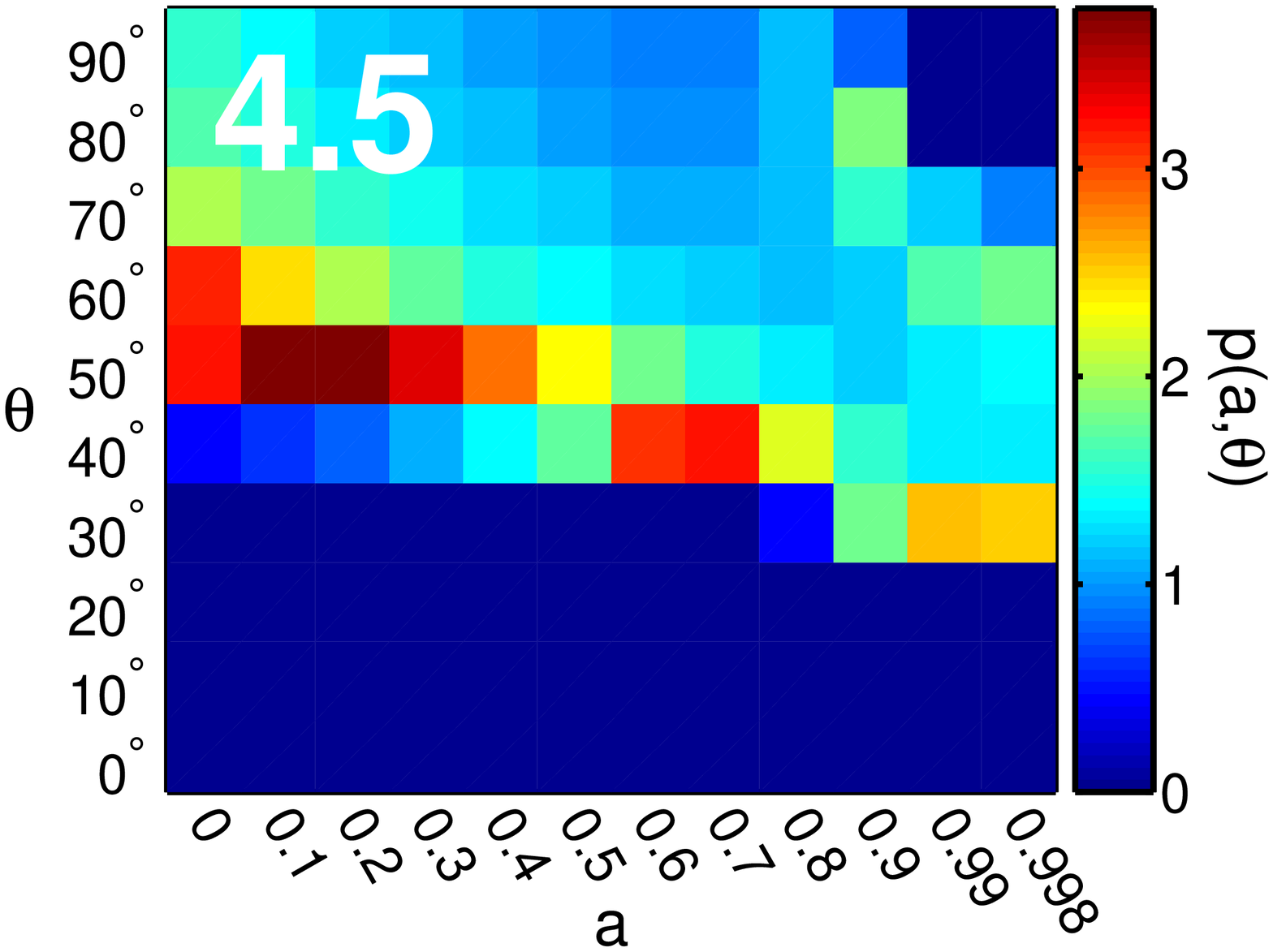}
\includegraphics[width=0.32\textwidth]{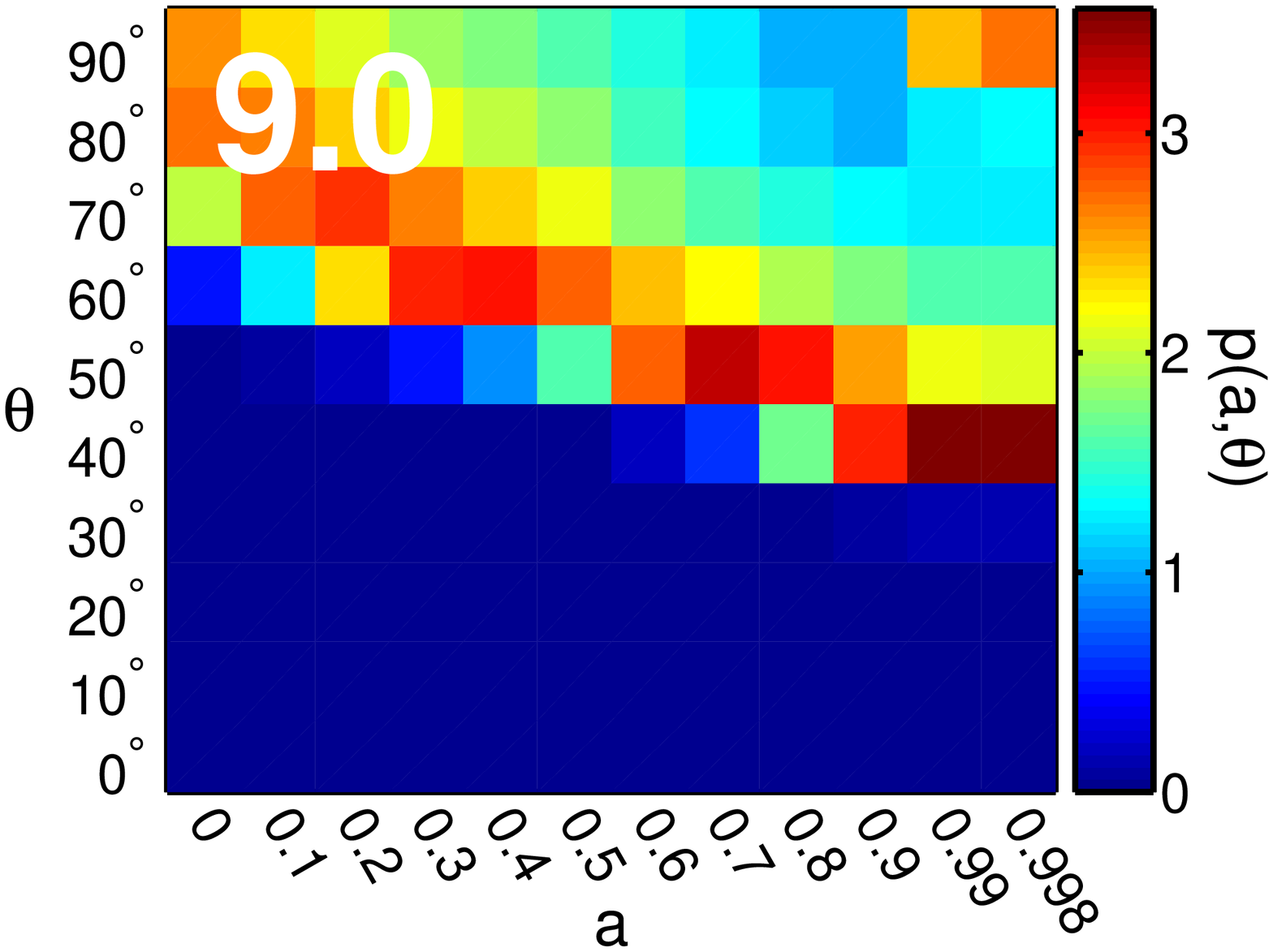}\\
\includegraphics[width=0.32\textwidth]{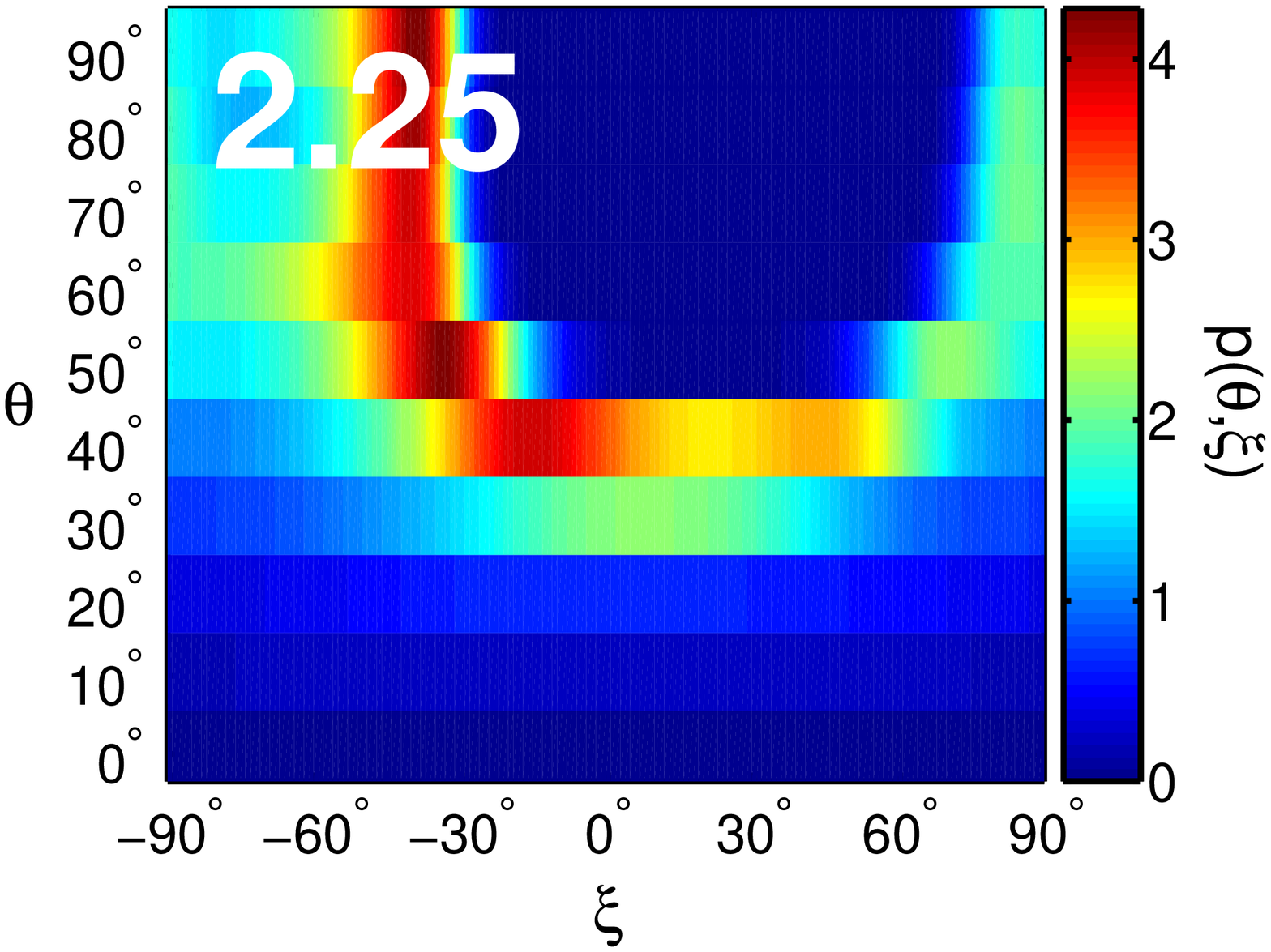}
\includegraphics[width=0.32\textwidth]{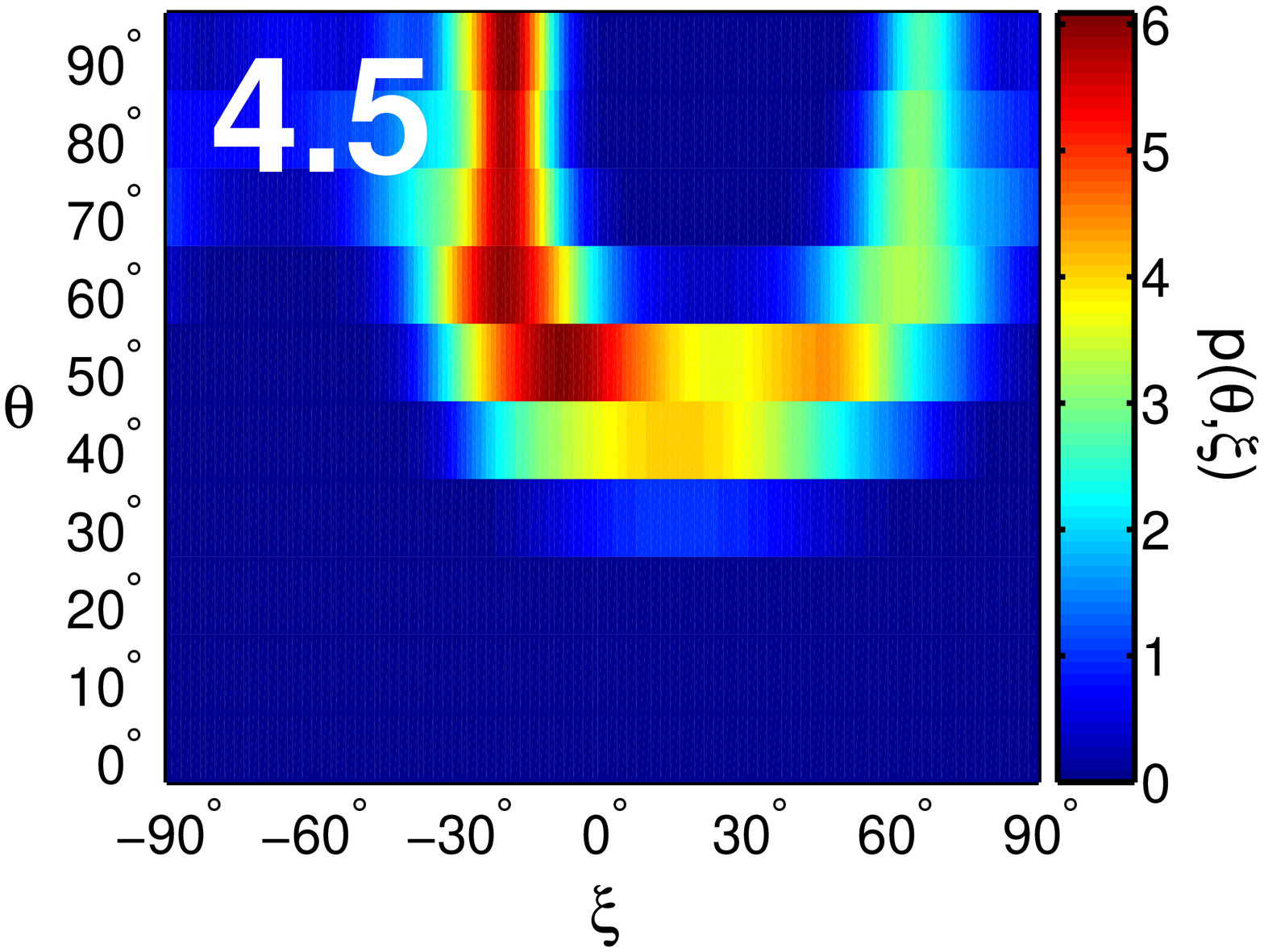}
\includegraphics[width=0.32\textwidth]{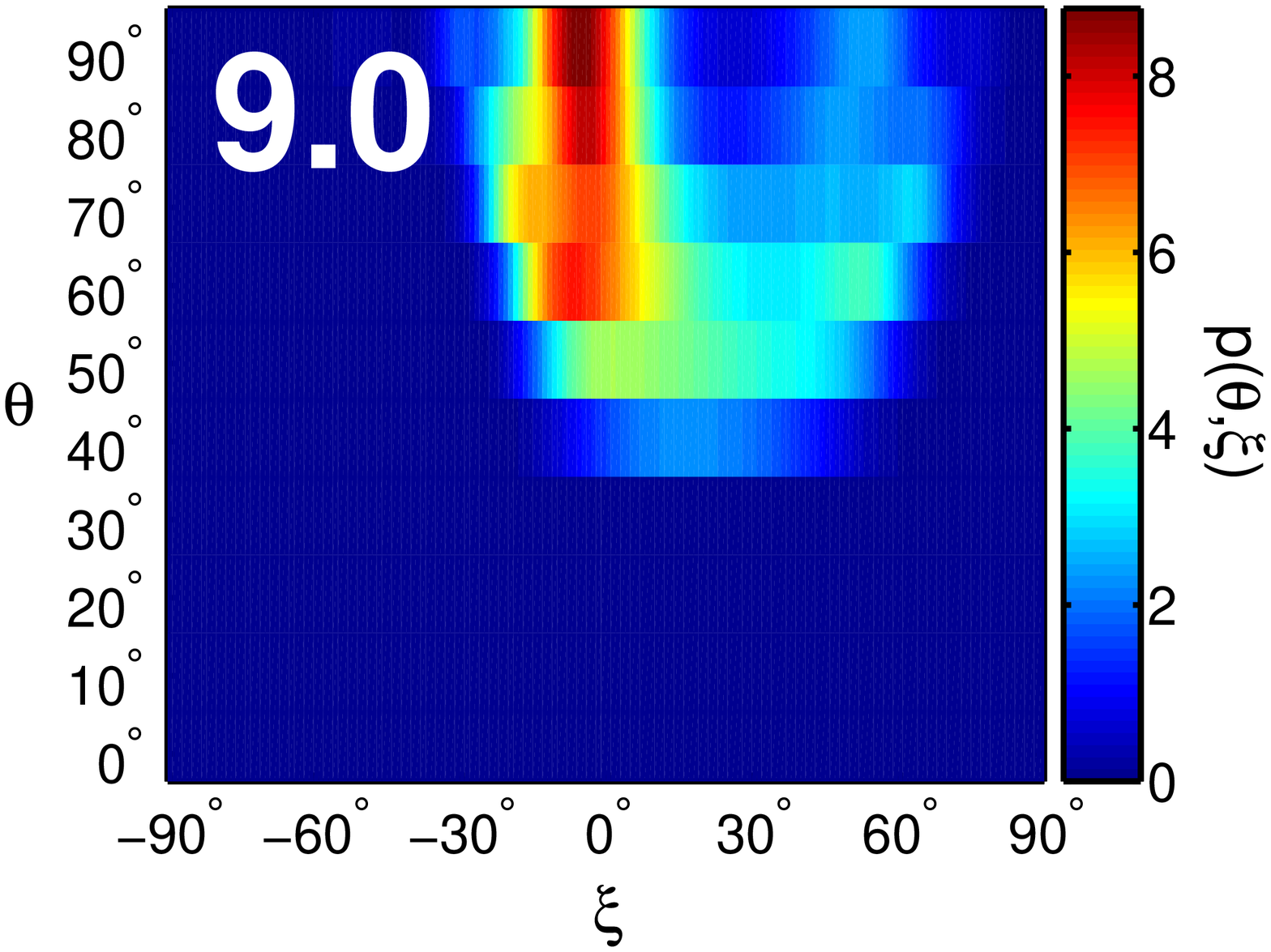}
\end{center}
\caption{{\em Top:} Probability densities of a given inclination and black
  hole spin assuming that $M_{\rm Sgr A*}=2.25$, $4.5$ and
  $9.0\times10^6\,M_\odot$ (note that the best current estimate is
  $4.5\pm0.4\times10^6\,M_\odot$).  {\em Bottom:} Probability
  densities of a given inclination and position angle for the same masses.  Note that the
  orientation of the accretion disk, and hence the black hole spin, is
  strongly constrained for all cases to lay in a taco-shell-shaped
  2-dimensional surface within the inclination-spin-position angle
  parameter space.  Regardless of mass, lower spins and moderate
  inclinations are preferred by the recent VLBI observations.  A
  position angle of $0^\circ$ corresponds to the projected spin vector
  being oriented North-South, and increases towards East.}\label{fig:mass}
\end{figure*}

\subsection{Visibility Modeling}
Visibilities are then defined in the normal way:
\begin{equation}
V(u,v) = \int\!\!\!\int \d x\, \d y \,\,\e^{-2\pi i (x u + y v)/\lambda} I(x,y)
\end{equation}
where $I(x,y)$ are the intensities on the image plane, with $x$
aligned East-West and $y$ North-South.  Because changing the position
angle, $\xi$ (vanishing at North and increasing towards the East),
corresponds simply to a coordinate rotation, the images were
originally computed only for a single position angle, namely $\xi=0$.
The visibilities associated with the ideal-resolution images were then
computed, and subsequently rotated to the desired $\xi$.  In practice,
the visibilities were calculated via a Fast Fourrier Transform, which
was padded sufficiently to resolve the shortest baseline
(\SMTO-\CARMA).  We also vary the total flux, or equivalently
$V_0\equiv V(0,0)$.  For sufficiently different values a new image
must be produced.  However, for small variations around our canonical
value of $2.4\,\Jy$ (within $0.5\,\Jy$) simply renormalizing the
visibilities produces a good approximation.  Finally, the
interstellar-scatter broadening was effected by multiplying the
visibilities in the $u$--$v$ plane by the Fourier-transformed
scattering kernel.

Since Sgr A* was detected on only two of the three VLBI stations,
there was insufficient observational data to determine the individual
baseline visibility phases.  Thus, henceforth, by visibility (and $V$)
we shall, more properly, refer to
the visibility magnitudes.  These are necessarily a function of
position in the $u$--$v$ plane, black hole spin ($a$), spin
inclination ($\theta$), position angle of the projected spin
vector ($\xi$) and flux normalization ($V_0$).  That is, for a
particular realization of the RIAF model, we have computed
$V(u,v;a,\theta,\xi,V_0)$. 

The ideal and broadened visibilities are shown in the center and
right-panels of Fig. \ref{fig:IVV}, respectively.  In both of these
the positions of the observed visibilities are shown by the white
points.  As expected, associated with the narrow-axis of the image crescent
is a correspondingly broad feature in the visibilities.  On the
maximum scale of interest (the \JCMT-\SMTO~baseline) the power is not
substantially reduced by the interstellar electron scattering.  At the
same time, along the long-axis of the image crescent the visibilities
drop off rapidly.  Thus, we may expect that, at the very least, the
VLBI observations will constrain $\xi$.  This turns out to be correct.

\section{Bayesian Data Analysis}\label{sec:DataAnalysis}

We follow a Bayesian scheme to compute the probability that given the
measured visibilities, a given set of model parameters
($a$,$\theta$,$\xi$) are correct.  This is, of course, predicated upon
the assumption that our simplistic RIAF model is appropriate for Sgr
A*.  Indeed, this is the primary uncertainty in our reported
constraints upon $a$, $\theta$ and $\xi$.  While we can address this
model dependence somewhat by comparing the results from different
black hole masses, at this point we can only hope that our results are
characteristic of generic RIAF models.  This is not unreasonable given
that the primary physics responsible for the structure of our images,
the Keplerian velocity profile, is a common theme among RIAF models.
Nevertheless, it remains to be proven.

Assuming the observational errors are Gaussian, the probability of
measuring a visibility $V_i$ {\em given} a particular set of model
parameters $(a,\theta,\xi,V_0)$ is
\begin{equation}
\begin{array}{l}
P_i(V_i|a,\theta,\xi,V_0)
= \nonumber\\
\displaystyle
\quad\frac{1}{\sqrt{2\pi}\Delta V_i}
\exp\left\{
- \frac{\left[V_i - V(u_i,v_i;a,\theta,\xi,V_0)\right]^2}{2\Delta V_i^2}
\right\}
d V_i
\,.
\end{array}
\end{equation}
This is appropriate for detections, i.e., along the \SMTO-\CARMA~ and
\JCMT-\SMTO~ baselines.  However, for the non-detection associated
with the \CARMA-\JCMT~ baseline this is not the case.  As discused in
Appendix \ref{app:ndp}, the probability of a non-detection given a
measurement threshold of $V_i$ and intrinsic uncertainty of $\Delta
V_i$ and expected value
$V(u_i,v_i;a,\theta,\xi,V_0)$ is
\begin{equation}
\begin{array}{l}
P_i(<V_i|a,\theta,\xi,V_0)
= \nonumber\\
\\
\displaystyle
\qquad\frac{1}{2}\left\{
1
+
\erf\left[
\frac{V_i-V(u_i,v_i;a,\theta,\xi,V_0)}{\sqrt{2} \Delta V_i}
\right] 
\right\}\,.
\end{array}
\end{equation}
Therefore, the probability of observing the measured set of
independent visibilities {\em given} a particular RIAF model is
\begin{eqnarray}
\displaystyle
P(\{V_i\}|a,\theta,\xi,V_0)
&=&
\displaystyle
\prod_{i=SC,\,JS}
P_i(V_i|a,\theta,\xi,V_0) \nonumber\\
&&
\displaystyle
\times
\prod_{j=CJ}
P_j(<V_j|a,\theta,\xi,V_0)\,,
\label{eq:Pvp}
\end{eqnarray}
where the first product is over the detections on the \SMTO-\CARMA~
({\em SC}) and \JCMT-\SMTO~ ({\em JS}) baselines, and the second is
over the non-detections on the \CARMA-\JCMT~ ({\em CJ}).

In order to assess the quality of the modeling at each spin and
inclination, we appeal to the Bayesian Information Criterion (${\rm BIC}$).
Specifically,
\begin{equation}
{\rm BIC} = -2 \ln\bigg[ P(\{V_i\}|a,\theta,\xi,V_0) \bigg] + k \ln n
\end{equation}
where $k=4$ is the number of parameters and $n=20$ is the number of data
points.  In the absence of the non-detection, this reduces to the
normal definition of $\chi^2$, up to an additive constant, and thus
comparing models with similar numbers of parameters reduces to the
normal $\chi^2$ minimization.  Since this latter statistic is commonly
used, and its properties generally well known, solely for the purpose
of assessing the quality of the fits we define an effective
$\chi^2$:
\begin{eqnarray}
\displaystyle
\chi^2
&=&
\displaystyle
-2 \ln\bigg[ P(\{V_i\}|a,\theta,\xi,V_0) \bigg]
+
\ln \left( \prod_i 2\pi \Delta V_i^2 \right)\\ \nonumber
&=&
\displaystyle
{\rm BIC} + {\rm const.}
\,,
\end{eqnarray}
Note that since only one of the visibility measurements is an upper
limit, we may expect that this definition of $\chi^2$ will behave very
similarly to the standard definition in our case.  The reduced
effective $\chi^2$ is shown as a function of spin and inclination in
Fig. \ref{fig:csq} for the best fit $V_0$ and $\xi$.  For nearly all
inclinations above $35^\circ$ a good fit can be found (indeed,
$\chi^2/\nu\simeq0.4$!).  The exception is at very high spins and high
inclinations (where the disk is edge on)\footnote{Note that in this procedure
we have ignored the statistics associated with fitting the spectrum,
choosing to break up these two components of the fitting process to
stress the implications of the new millimeter VLBI observations.}.

While eq. (\ref{eq:Pvp}) gives the probability that the observed visibilities come from
a given model, what we would like to know is somewhat different: the
probability of a set of model parameters given the observed visibilities.
That is, we would like the probability density
$p(a,\theta,\xi,V_0|\{V_i\})$.  With an appropriate choice of priors
on $a$, $\theta$, $\xi$ and $V_0$, we may construct the desired
probabilities via Bayes' theorem.  As such, we now turn to the
problem of choosing these priors.

A natural choice for the prior upon $\theta$ and $\xi$ comes from the
assumption that Sgr A*'s spin orientation probability is isotropic,
i.e., we have no other information regarding its direction.  This
results in $\wp (\theta,\xi) = \sin\theta$.  In the absence of a complete
theoretical understanding of the spin evolution of supermassive black
holes, we choose the prior on $a$ to be uniform, i.e. $\wp(a)=1$.
Finally, we set the prior on $V_0$ to be uniform as well.  Since the
allowed range of variation in $V_0$ is small, and the prior
probability is expected to vary smoothly, this is not a significant
oversight.  Therefore, Bayes' theorem gives
\begin{equation}
\begin{array}{l}
\displaystyle
p(a,\theta,\xi,V_0|\{V_i\})
\nonumber\\
\\
\qquad\displaystyle
=
\frac{\displaystyle
P(\{V_i\}|a,\theta,\xi,V_0) \wp(a) \wp(\theta,\xi)\wp(V_0)
}{\displaystyle
\int\d a  \d \theta\d \xi \d V_0 \, P(\{V_i\}|a,\theta,\xi,V_0) \wp(a) \wp(\theta,\xi) \wp(V_0)
}
\nonumber\\
\\
\qquad\displaystyle
=
\frac{\displaystyle
P(\{V_i\}|a,\theta,\xi,V_0) \sin\theta
}{\displaystyle
\int\d a  \d \theta\d \xi \d V_0 \, P(\{V_i\}|a,\theta,\xi,V_0) \sin\theta
}\,.
\end{array}
\end{equation}

\begin{figure*}[th!]
\begin{center}
\includegraphics[width=\textwidth]{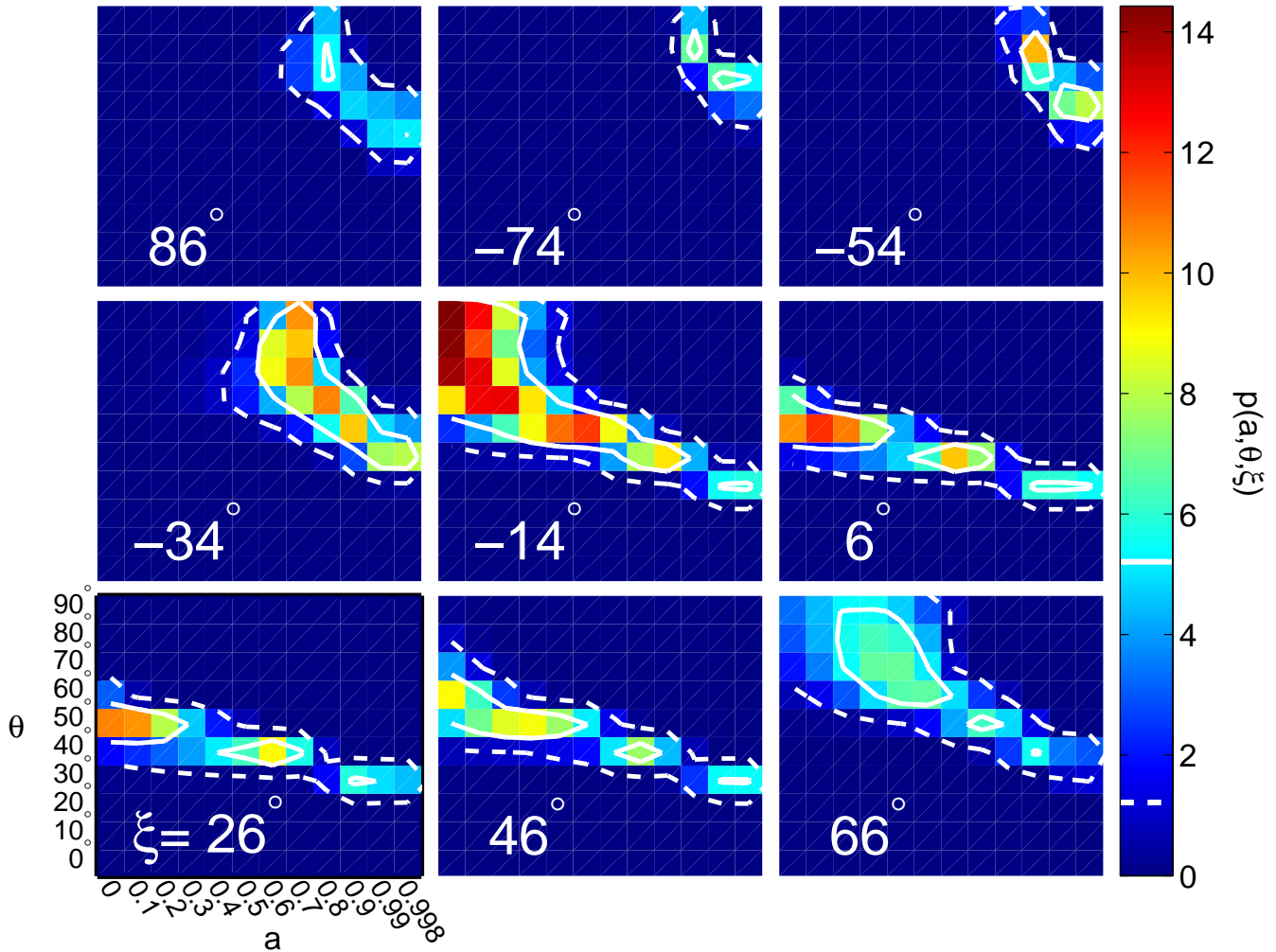}
\end{center}
\caption{$p(a,\theta,\xi)$ is shown for slices of constant
  $\xi$.  The slice with the maximum over-all probability is shown in
  the center panel.  The white solid and dashed lines show the
  $1$-$\sigma$ and $2$-$\sigma$ contours, defined explicitly in the
  text, respectively.  Note that while the probability is quite peaked, likely
  values of the parameters exist for a large number of spins,
  inclinations and position angles.}\label{fig:slices3D}
\end{figure*}

\begin{figure*}[t!]
\begin{center}
\includegraphics[width=0.33\textwidth]{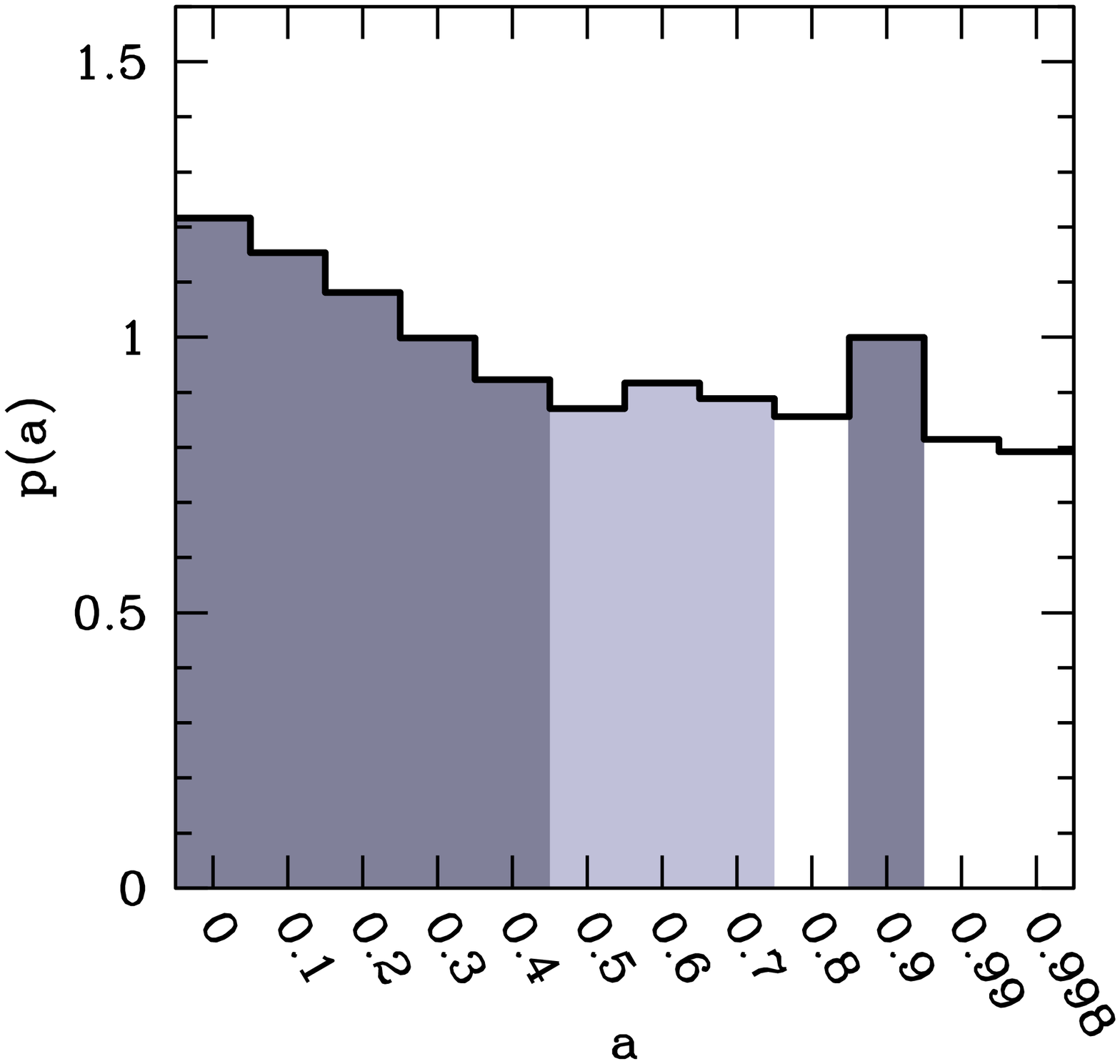}
\includegraphics[width=0.33\textwidth]{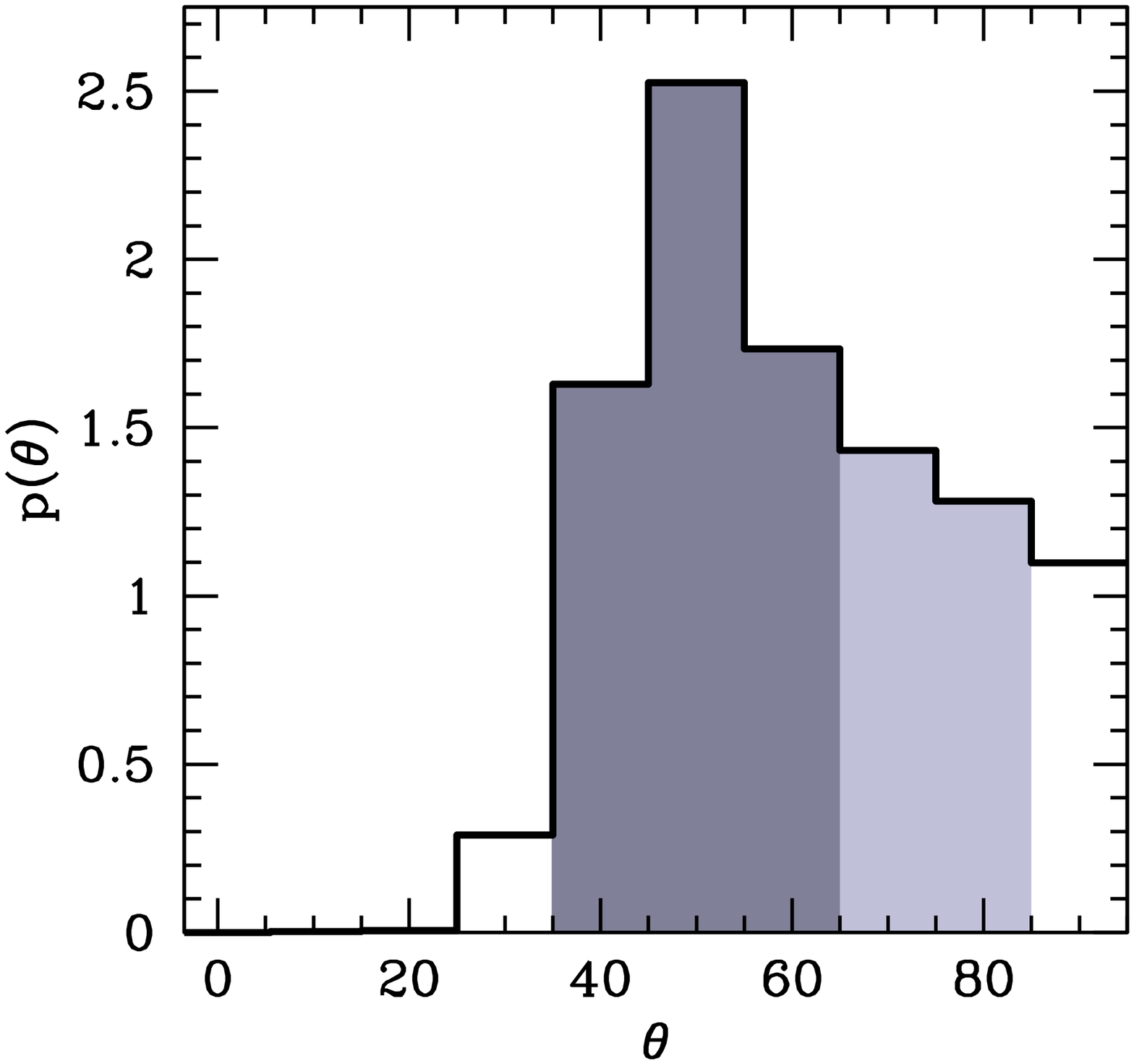}
\includegraphics[width=0.33\textwidth]{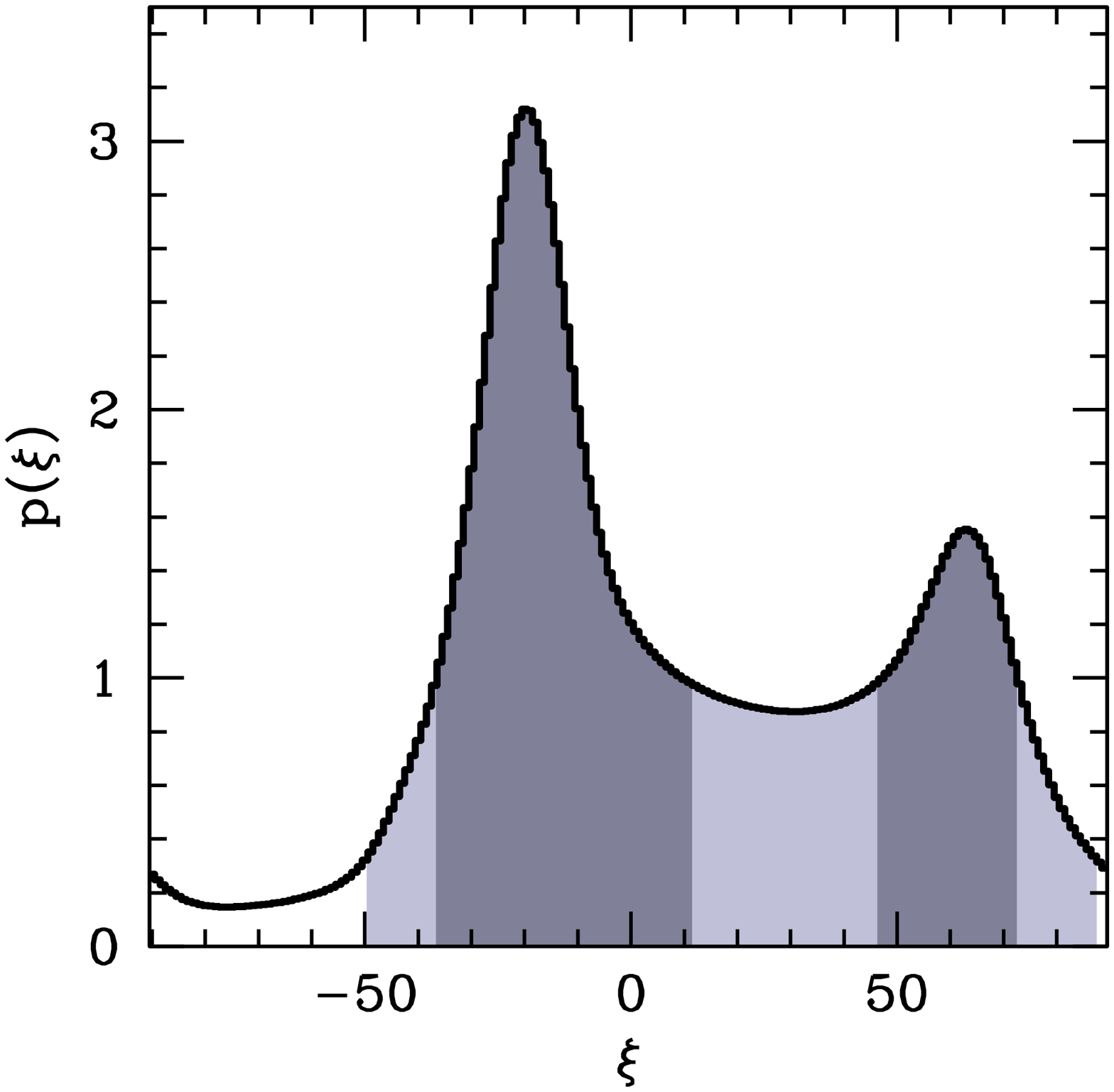}
\end{center}
\caption{From left to right: $p(a)$, $p(\theta)$ and $p(\xi)$, all
  marginalized over all other parameters.  In all cases the
  probability distribution is highly non-Gaussian.  The dark and light
  shadded regions denote the $1$-$\sigma$ and $2$-$\sigma$ regions, as
  defined in the text, respectively.  It important to keep in mind
  that parameter estimates from these distributions are strongly
  correlated.}\label{fig:marg1D}
\end{figure*}

This is necessarily a probability density in a four-dimensional
parameter space, and thus is quite difficult to visualize directly.
Furthermore, some of these parameters are physically more interesting
than others.  Therefore, we construct a variety of marginalized
probabilities from this for presentation and analysis.  The most
general is simply marginalized over $V_0$:
\begin{equation}
p(a,\theta,\xi) = \int \d V_0\, p(a,\theta,\xi,V_0|\{V_i\})\,.
\end{equation}
This probability distribution is shown explicitly in
Fig. \ref{fig:slices3D}.  We construct a pair of two-dimensional
marginalized probability densities as well:
\begin{equation}
p(a,\theta) = \int\d\xi \d V_0\, p(a,\theta,\xi,V_0|\{V_i\})
\label{eq:marg_xi}
\end{equation}
and
\begin{equation}
p(\theta,\xi) = \int\d a \d V_0\, p(a,\theta,\xi,V_0|\{V_i\})
\label{eq:marg_a}
\end{equation}
These are plotted in Fig.'s \ref{fig:marg} and \ref{fig:mass}.
Alternatively, we could choose the most likely values of either $V_0$
or $\xi$.  For some of the panels in Fig. \ref{fig:marg}, we choose a
hybrid probability: marginalized over $\xi$ but the most likely in
$V_0$.  Finally, for the purpose of identifying the probability
distribution of each parameter separately, we also construct the
marginalized one-dimensional probability densities:
\begin{eqnarray}
p(a) &= \displaystyle \int \d\theta \d\xi \d V_0\, p(a,\theta,\xi,V_0|\{V_i\})\\
p(\theta) &= \displaystyle \int \d a \d\xi \d V_0\, p(a,\theta,\xi,V_0|\{V_i\})\\
p(\xi) &= \displaystyle \int \d a \d\theta \d V_0\, p(a,\theta,\xi,V_0|\{V_i\})\,.
\label{eq:marg1_a}
\end{eqnarray}
These are shown in Fig \ref{fig:marg1D}.

\section{The Nature of the VLBI Constraints}\label{sec:Results}

The relative importance of the three different baselines
is shown in Fig. \ref{fig:marg}.  To make a direct comparison of the
contributions from different baselines possible, the top-left,
top-right and bottom-left panels of Fig. \ref{fig:marg} show the
probability after setting $V_0$ to the most likely value and
marginalizing over $\xi$ (justified by the fact that the probability
distributions in $\xi$ are quite similar, while those in $V_0$ were
not).  Both the \SMTO-\CARMA~detections and the \CARMA-\JCMT~upper
limit exclude the high-spin, high-inclination portion of the parameter
space.  This is primarily because high-spin RIAF models result in
images that are sufficiently compact to substantially over-predict the
flux observed on both of these baselines.  Due to its considerably
longer baseline length, the \CARMA-\JCMT~non-detection is more
constraining than the multiple \SMTO-\CARMA~detections, despite only
representing an upper-limit.

As anticipated, most constraining are the visibilities measured on the
\JCMT-\SMTO~baseline.  This baseline convincingly excludes low
inclinations (face-on disks).  This may be easily understood,
qualitatively, in terms of the RIAF images themselves.  As inclination
increases, the RIAF image becomes more asymmetric, and increasingly
dominated by a thin crescent (e.g., left-panel of Fig. \ref{fig:IVV}).
It is generally the direction across the minor-axis of the
crescent that has the shortest scale intensity variations, and
consequently determines the long-baseline visibilities.  Below a
critical inclination, the crescent grows sufficiently fat that it is
resolved out by the \JCMT-\SMTO~baseline.

At large inclinations, the crescent becomes too thin, requiring the
\JCMT-\SMTO~baseline to be oriented obliquely relative to the spin
axis of Sgr A*.  This restricts the
available values of $\xi$ to an increasingly limited range, as may be
seen explicitly in the bottom-center panel of Fig. \ref{fig:mass}.
Thus, while it is nearly always possible to obtain a satisfactory fit,
the reduced range in position angle makes such a configuration
unlikely.  Hence we find that the long-baseline visibilities strongly
constrain the possible values of inclination and spin to a narrow band
near $\theta\simeq50^\circ$ and generally favoring moderate $a$.  We
note that the appearance of islands of high probability in the
top-right panel of Fig. \ref{fig:marg} is almost certainly due to
under sampling in $\theta$ near this critical region, and not
indicative of a bifurcation in the allowed parameter space.

This behavior is clearly visible in the combined $p(a,\theta)$, which
closely resembles that from the \JCMT-\SMTO~baseline alone.  In this
case, we have marginalized over $V_0$ as well as $\xi$ (i.e.,
$p(a,\theta)$ as defined by eq. \ref{eq:marg_xi}), though the
difference had we chosen the most likely value of $V_0$ is less than
10\% everywhere.  The shorter baselines have effectively
removed a small high-spin, high-inclination island that persisted
in the \JCMT-\SMTO~probability distribution alone.  Additionally, they
have further restricted the spin to low-to-moderate levels.  The
combined result, however, is to restrict the spin and inclination to a
narrow strip.

In a similar fashion, $p(\theta,\xi)$ appears to limit RIAF models to
a narrow band of orientations.  Due to the approximate up-down
symmetry of the image (parallel to the projected spin axis), the
corresponding band has a characteristic ``U'' shape, corresponding to
something akin to a taco shell in the $a$,$\theta$,$\xi$ parameter
space.  (Note that we do not plot $\xi$ over the entire $360^\circ$
range due to the symmetry of the visibilities under reflection.)

In order to ascertain how the present uncertainty in the mass of Sgr
A* effects our conclusions, we repeated the analysis for black hole
masses of $4.1\times10^6\,\Ms$ and $4.9\times10^6\,\Ms$.  Since the
mass uncertainty is strongly correlated with the distance to Sgr A*
($M D^{-1.8}$ is very well determined by stellar orbits, \citealt{ghez09,gillessen08}), we
altered the distance to the Galactic center accordingly.  As a
consequence, the roughly 10\% change in the mass results in a 4\%
change in the angular scale of the RIAF images.  A correspondingly
small change was seen in the resulting probability distributions,
implying that, apart from the the RIAF modeling, the paucity of
millimeter-VLBI observations is the dominant obstacle to constraining
the black hole spin properties.

Altering the angular scale of the images also provides a proxy, albeit
a poor one, for considering different Sgr A* models, corresponding to
different disk scale lengths.  $p(a,\theta)$ and $p(\theta,\xi)$ for a
black holes of mass $2.25\times10^6\,\Ms$ and $9.0\times10^6\,\Ms$ are
compared to those determined using the estimated value of
$4.5\times10^6\,\Ms$ in Fig. \ref{fig:mass}.  These extreme mass
changes correspond to a 30\% change in the angular scales of the
images, and can make a substantial difference to the resulting
probability distributions.  Nevertheless, high inclinations and low
spins are still generally preferred, though less so for large images.

Finally, because Sgr A* is an inherently dynamical environment,
exhibiting substantial variations in the millimeter flux on $30\,\min$
time scales, we may not be justified in assuming that the image of Sgr
A* was stationary during the entire time that observations were made.
Indeed, searching for variations in the VLBI closure quantities has
been suggested as a way to directly probe the existence of hot spots
in Sgr A*'s accretion flow \citep{doeleman08b}.  Therefore,
to check this, we repeated this analysis for the two days over which
the observations were performed separately, finding no significant
difference.  This implies that either Sgr A* was quiescent during this
time or that the individual $3.5\,{\rm hr}$ observation windows were
sufficiently long to average out this activity.  The anomalously low
$1.3\,\mm$ flux suggests the former interpretation is correct.

\section{Parameter Estimation}\label{sec:parameters}

It is evident from the previous section that the allowed parameter
space is highly non-Gaussian.  As such, we must take special care in
how we extract values and their attendent uncertainties for the
fitting parmeters.  In all cases, these values will be highly
correlated, and the systematic uncertainties due to the choice of a
particular RIAF model will dominate the errors (and thus will not be
reviewed again in this section).

In order to determine $1$-$\sigma$ and $2$-$\sigma$ error surfaces, we
first define a cumulative probability
\begin{equation}
P(>p) = \int_{p(\bmath{x})\ge p} p(\bmath{x}) \, \d \bmath{x}
\end{equation}
where $\bmath{x}$ are the parameters of the total probability
distribution.  That is, $P(>p)$ is simply the probability associated
with the region of the parameter space that has probability density
above $p$.  This is necessarily a monotonic function of $p$, which may
then be inverted to find $p$ as a function of $P(>p)$.  We then define
the $1$-$\sigma$ contour to be that associated with the $p$ for which
$P(>p)=0.683$. Similarly, we define the $2$-$\sigma$ contour by
$P(>p)=0.954$.  While these satisfy the normal definition of
$1$-$\sigma$ and $2$-$\sigma$ errors, it is important to remember that
in our case the errors are strongly non-Gaussian.

Fig. \ref{fig:slices3D} shows the probability density in the full
three-dimensional parameter space as a sequence of constant-$\xi$
slices.  The most likely parameter combination is $a=0^{+0.2+0.4}$,
$\theta={90^\circ}_{-40^\circ-50^\circ}$ and $\xi={-14^\circ}^{+7^\circ+11^\circ}_{-7^\circ-11^\circ}$.  However, as remarked in the
previous section, there are acceptable solutions for a wide range of
$a$, $\theta$ and $\xi$.  The primary effect of the VLBI measurements
is to restrict these parameters to a band around a two-dimensional
surface, resembling a taco shell with its sides around
$\xi\simeq-20^\circ$ and $\xi\simeq63^\circ$, the highest probability
densities clustered on the former.  Generally, we see a preference for
low spins and can rule out inclinations below $30^\circ$.

More diagnostic for single parameters are the fully marginalized
probability distributions shown in Fig. \ref{fig:marg1D}.  It must be
remembered, however, that these parameter estimates are strongly
correlated, and thus these estimates should be used with caution.
Again we use the cumulative probability, $P(>p)$ to define the
$1$-$\sigma$ and $2$-$\sigma$ surfaces, shown in the panels of
Fig. \ref{fig:marg1D} as the dark and light shaded regions,
respectively.

From the left panel  of Fig. \ref{fig:marg1D}, the most likely spin is
$a=0^{+0.4+0.7}$. While we can rule out very high spins ($a\ge0.99$)
at the $2$-$\sigma$ level for our particular RIAF model, the spin is
otherwise weakly constrained.  Moreover, it appears to have a small
high-spin island, at the $1$-$\sigma$ level, around $a=0.9$.
The presence of this island is limited by the
\CARMA-\JCMT~non-detection, and therefore can be directly probed via
future observations.

In contrast, the inclination is fairly robustly limited.  The most
likely inclination is
$\theta={50^\circ}^{+10^\circ +30^\circ}_{-10^\circ-10^\circ}$, where
the lopsided errors are due to the lopsided nature of the probability
distribution.  It is very clear in the center panel of
Fig. \ref{fig:marg1D} that face-on geometries  ($\theta\le 30^\circ$)
are convincingly ruled out.  However, there is a tail extending to
higher inclinations\footnote{Indeed, as mentioned above, the most likely
overall set of parameters has $\theta=90^\circ$.}.

The distribution of position angles is more complicated than either
spin or inclination.  In the right panel of Fig. \ref{fig:marg1D} we
see the taco-shell geometry most clearly in terms of the bimodal
distribution of likely $\xi$.  The most probable position angle is
$\xi={-20^\circ}^{+31^\circ+107^\circ}_{-16^\circ-29^\circ}$.
However, a second solution does exist at
$\xi={63^\circ}^{+9^\circ+24^\circ}_{-17^\circ-112^\circ}$, though
containing only $38\%$ of the probability (under the $1$-$\sigma$
peak) of the more likely solution.

Our limits upon the inclination are in quite good agreement with
previous efforts.  \citet{huang07} fit a qualitatively different RIAF
model, primarily neglecting orbital motion, at $3.5\,\mm$ and
$7\,\mm$.  While these favor high inclinations, the fact that the
images are dominated by interstellar scattering prevents a conclusive
measurement.  Similarly, efforts to fit Sgr A*'s flares with
hydrodynamic instabilities find an inclination of roughly
$77^\circ\pm10^\circ$, though the systematic uncertainty associated
with their model is unclear.  Fitting a qualitatively different model to the
long-wavelength observations, a hydrodynamic jet, \citet{markoff07}
also favor large inclinations ($\theta\gtrsim 75^\circ$).  However, it
is not obvious how this qualitative agreement should be interpreted, given the
very different geometries involved.

Unlike the inclination, our position angle estimate disagrees
significantly with many previous efforts.  The primary solution is in
good agreement with the $3.5\,\mm$ and $7\,\mm$ fits of RIAF models by
\citet{huang07}, which imply $-50^\circ\lesssim \xi \lesssim 30^\circ$.  
While there is considerable uncertainty in their 
estimates (like ours), we note that their result is not consistent
with our second solution at the 1-$\sigma$ level.  In contrast, only
our second solution is in agreement with analyses of NIR polarization
observations, assuming a simplistic orbiting hot-spot model, which
find $60^\circ\lesssim\xi\lesssim105^\circ$ \citep{meyer07}.  This is
not particularly surprising given the difficulties in assessing the
relationship between the emitted polarization and the underlying
plasma geometry.  At the 1-$\sigma$ level, none of our solutions are
consistent with efforts to fit jet models to Sgr A* at $7\,\mm$,
which find $80^\circ\lesssim\xi\lesssim120^\circ$ \citep{markoff07}.
This is also not unexpected given the considerably different
geometries involved.  Between all of these (even prior to our
estimate), nearly all values of $\xi$ have been reported, suggesting
that significant improvements in theoretical understanding and
additional observations are required before this question can be
settled.

Finally, note that our estimates imply that the X-ray feature
described in \citet{muno08} is not associated with a putative jet, but
simply another filamentary structure in the vicinity of Sgr A*.  However,
interestingly, our primary position angle estimate
\citep[and that of][]{huang07} is similar to the orientation of the
compact, clockwise disk of massive stars orbiting the Galactic center
\citep{genzel03}.  This is somewhat surprising given the the large
radii ($2\times10^5 GM/c^2$) of the stellar disk, though may be
natural if the observed stars are the remnants of an active period in
Sgr A*'s recent past.

\section{Conclusions}\label{sec:Conclusions}

Despite the sparse $u$--$v$ coverage, and the existence of a detection
on only one long baseline, we have been able to significantly
constrain the possible parameters of an accretion flow onto Sgr A*.
Within the context of a qualitative RIAF model that fits the observed
spectra of Sgr A*, we have have been able to substantially constrain the
orientation of Sgr A*'s spin.  The magnitude of this spin is less well
determined, though the black hole cannot be maximally rotating.  This
result is relatively insensitive to the black hole mass.

We have not ascertained the strength of our constraints' dependence
upon the particular RIAF model, though our estimates are consistent
with earlier efforts comparing longer wavelength observations to
an alternative RIAF model.  However, there are two additional reasons
to believe that our results will be generic of RIAF models.  The
first is that the underlying physics that limits the orientation is
the Doppler beaming and boosting that is dependent primarily upon the
Keplerian velocity profile of the accretion flow, a feature that is
generic among RIAF models.  The second is the weak dependence
upon large variations in mass, implying that changing the scale
lengths of the accretion model does little to alter our results.
Thus, we expect the qualitative form of our constraints to be a
generic feature of all RIAF models for Sgr A*, and our quantitative
results to be roughly correct.  Of course, should the emission
observed from Sgr A* not arise in an accretion flow, our results could
be quite different.

As implied by the right-panel of Fig. \ref{fig:IVV}, additional
long-baseline observations are sorely needed to unambiguously
determine both the applicability of RIAFs to Sgr A* and constrain
their parameters.  In a companion paper, \citet{fish08} we report
upon an analysis of the ability of possible millimeter VLBI arrays to
do so.  Given the number of potential radio telescopes that could be
added to the current $1.3\,\mm$ VLBI array \citep{doeleman08b}, the
prospects for making significant progress in model parameter
estimation are excellent.  Already it is clear that
with the advent of millimeter VLBI we are now entering an era of
precision black-hole accretion physics.

\acknowledgments

\appendix

\section{Non-detection Probability} \label{app:ndp}
Consider an observable $y$, with an expected value of $y_0$ and
Gaussian random errors of amplitude $\Delta y$.  Then, the probability
density of measuring a value $y$ is
\begin{equation}
p(y) = \frac{\e^{-(y-y_0)^2/2\Delta y^2}}{\sqrt{2\pi}\Delta y}\,.
\end{equation}
A non-detection, by definition, corresponds to a ``measured'' value of
$y$ below some threshold $Y$.  The probability of such an event is
simply
\begin{equation}
P(<Y)
=
\int_{-\infty}^Y p(y) \d y
=
\frac{1}{2}
+
\int_{y_0}^Y \frac{\e^{-(y-y_0)^2/2\Delta y^2}}{\sqrt{2\pi}\Delta y} \d y
=
\frac{1}{2} \left[
1
+
\erf\left( \frac{Y-y_0}{\sqrt{2}\Delta y} \right)
\right]\,,
\end{equation}
where
\begin{equation}
\erf(x) \equiv \sqrt{\frac{2}{\pi}} \int_0^x \e^{-t^2} \d t\,,
\end{equation}
is the standard error-function.  As expected, when $Y=y_0$,
$p(<y_0)=1/2$ and when $Y\gg y_0$, $p(<y_0)\simeq 1$.  This procedure
is identical to that employed by \citet{kelly07}, \S5.2, and
references therein.  While in that work the primary interest was a
careful investigation of the ability of Bayesian techniques in the
context of linear regression, we are faced with a more general fitting
problem.  Nevertheless, how non-detections, or ``censored data'', enter
into the likelihood is identical, the distinctions arising only later
in the analysis.

\end{document}